%% file: ms.tex
\documentclass[12pt,preprint]{aastex}

\shortauthors{Chiboucas \& Mateo}
\shorttitle{Centaurus Cluster Galaxy Radial Velocities}

\begin{document}

\title{The Luminosity Function of Nearby Galaxy Clusters II: Redshifts and Luminosity Function for Galaxies in the Region of the Centaurus Cluster}

\author{Kristin Chiboucas\altaffilmark{1}}
\affil{Gemini Observatory}
\affil{670 N. A'ohoku Pl, Hilo, HI 96720-2700} 
\email{kchibouc@gemini.edu}
\altaffiltext{1}{Work done while at The University of Michigan}

\and \author{Mario Mateo}
\affil{University of Michigan}
\affil{Department of Astronomy, Univ. of Michigan, Ann Arbor, MI 48109-1090}
\email{mmateo@umich.edu}

\begin{abstract}

We acquired spectra for a random sample of galaxies within a 0.83 square degree 
region centered on 
the core of the Centaurus cluster. Radial velocities were obtained for 225 galaxies
to limiting magnitudes of $V < 19.5$.  Of the galaxies for which velocities
were obtained, we find 35\% to be member galaxies.  New redshifts are obtained
for 15 Centaurus cluster members, many of these dwarf galaxies.  
Radial velocities for the other members
agree well with those from previous studies.  
Of the 78 member galaxies, magnitudes range from $11.8 < V < 18.5$
($ -21.6 < M_{V} < -14.9$ for H$_{\circ} = 70$ km s$^{-1}$ Mpc$^{-1}$) with a limiting
central surface brightness of $\mu_{\circ} < 22.5$ mag arcsec$^{-2}$.
While many of these galaxies are giants, about 25 galaxies with
M$_{V} > -17.0$ are considered dwarfs.
We constructed the cluster galaxy luminosity function by using 
these spectroscopic
results to calculate the expected fraction of cluster members in each magnitude
bin. The faint-end slope of the luminosity function using this method is 
shallower than the one obtained using a statistical 
method to correct for background galaxy contamination.  
We also use the spectroscopy results
to define surface brightness criteria to establish membership for
the full sample.  
Using these criteria, we find a luminosity function 
very similar to the one constructed with the statistical background correction.
For both, we find a faint-end slope $\alpha \sim -1.4$.  The error in faint-end
slope for the statistically corrected LF is $\sim\pm0.2$.  Adjusting the
surface brightness membership criteria we find that the data are 
consistent with a faint-end slope as shallow as $-1.22$ or as steep as $-1.50$.
We describe in this paper some of the limitations of using these 
methods for constructing the galaxy luminosity function. 
This is paper II in our investigation of the cluster galaxy luminosity
function.

\end{abstract}


\section{Introduction\label{introsec}}

There is a 
paucity of spectroscopic and kinematical information on dwarf galaxies
in clusters.  Most existing data comes from the nearby Virgo and 
Fornax clusters (see e.g. \citet{cgw01}, \citet{djgp00}, \citet{pgcsg02},
\citet{ggm03}, \citet{zsh04}, \citet{gzsbb04}).
Data also exist for Centaurus (\citet{sjf97} and this work) and 
Coma (\citet{ecbcmp02} and references therein).
Even within these relatively nearby clusters, spectroscopy of dwarf
galaxies is a challenging task, requiring spectra for (typically)
low-surface brightness galaxies with integrated magnitudes as faint as
$R \sim 20$, or $B \sim 21$.  

Nonetheless, obtaining large kinematic samples of cluster dwarf galaxies
is important for a number of reasons.
Processes involved in cluster formation and evolution
leave dynamical signatures on the velocity distribution
of member galaxies.
A comparison of the velocity distributions along with spatial distributions
as a function of galaxy type provides important clues about the
dynamical history of the clusters and of the different galaxy populations
\citep{ecbcmp02}.
Several possible formation mechanisms for
dwarf galaxies may be tested in this way.  Dwarfs may have formed, for 
example, early
on in the cluster, later from infalling spirals which are subjected to
stripping/harassment, from the transformation of infalling dIrrs, or may
instead originate from
a population of satellites previously bound to infalling spirals.  
The velocity distribution and dispersion coupled with spatial 
distributions 
will indicate whether the population is highly clustered and virialized, or
has a larger spread indicative of infall or substructure. 
The few studies of cluster kinematics in Fornax, Virgo, Coma, and Centaurus 
have revealed evidence for
a number of these formation mechanisms and hint at different
evolutionary scenarios for the clusters \citep{djgp00,cgw01,ecbcmp02,sjf97}.

At a more fundamental level, kinematic studies of clusters help
establish cluster membership independent of and complementary to
other methods designed to determine background contamination.
Most studies of the cluster galaxy-luminosity function (LF), for example,
make use of dedicated control fields to statistically determine the background
contribution to the cluster fields.  
This approach assumes that the control field is representative of the
cluster's background population; the existence of significant cosmic
field-to-field variance in deep galaxy counts \citep{hjddgo00} partly
undermines this assumption and necessarily introduces large uncertainties
into the final luminosity function results.  Only for the
nearest clusters such as Virgo, Fornax, and Centaurus have 
morphological means been used to determine the membership
status of each galaxy (see \citet{jd97}).  When morphological information is
not available, or to supplement it, some studies make use of surface
brightness and color information \citep{shp97} to single out members.
None of these methods is entirely satisfactory, especially when they are used in isolation.
While the statistical method relies on assumptions and possible systematic
biases (which could be present if clusters are initially picked out, for example,
due to higher background densities), morphological or color selection
methods may miss entire populations of dwarf galaxies such as very
compact, M32-like galaxies \citep{dg98}.  
Only redshifts can reliably establish cluster membership.

Recent spectroscopic studies illustrate this point.  Using kinematic
observations to confirm membership of dwarf candidates,
\citet{anmdhl98} and \citet{shco98} have found that Coma contains
fewer dwarf members than expected based on the cluster LF determined
from statistical subtraction of background counts.
Using results from both studies, a combined
total of 5 out of 46 galaxies in the magnitude range $19 < R < 21.3$
were found to be members.   
Based on the Coma LF measured by \citet{bntuw95}, $16\pm11$ out of 46
are expected.  
The quoted error includes the poisson error contribution from both 
control and cluster counts.
Though the count errors in this case are large,
this example shows how kinematic observations can test
the precision of statistical background galaxy subtraction, especially
at the faint end of the LF.

This is particularly relevant in light of recent studies that question
the validity of measurements of deep cluster LFs.
\citet{vml01} propose that the steep faint-end slopes of LFs which
have been found in numerous studies \citep{dpdmd94,dphm95,t97b,sdp97,dal02} 
are due to
projection effects which cannot be accurately corrected for
by subtracting random fields.  In a subsequent paper, \citet{vmml03}
argue that Abell clusters, which were initially
picked out by eye simply as overdense regions,
suffer from background contamination so severe that their cluster
status may largely be due to projection effects. One implication is that the 
faint-end slopes of the LFs of Abell clusters will be systematically 
overestimated.  
Supporting evidence comes from clusters detected in X-ray which have LFs that 
are typically flat, while
non X-ray clusters have much steeper faint-end slopes.  Since they
find no correlation of LF slope with X-ray brightness, they
argue that the non X-ray clusters are merely the superposition of
multiple poor groups or spurious unbound systems.  When background 
contamination is corrected using random control fields, with
presumably fewer background structures, these galaxy counts are
undercorrected leaving an excess of galaxies assumed incorrectly to be part
of a large cluster population.  
Since this interpretation questions the accuracy of our census of even
local cluster populations, it is important to determine if this is a
general problem, and to understand its effects on the galaxy LFs of
individual clusters.  Kinematic observations are the best way to help
address this issue.

The Centaurus Cluster is a nearby (z=0.0114) rich and X-ray bright cluster 
of galaxies.
\citet{dcl86} were the first to perform a comprehensive study
of the Centaurus cluster.  They published a catalog of 319 predominantly
bright ($G < 17.5$) galaxies including radial velocities for 259 
of these.  
From the redshifts of these bright galaxies, they discovered that 
the Centaurus cluster exhibits a bimodal velocity distribution.  
These data suggest that a smaller group centered at 4500 km/s with low velocity 
dispersion around 260 km/s is infalling into
the main body of the cluster, which is centered at around 3000 km/s
and has a large velocity dispersion of 933 km/s.  

More recently, \citet{jd97} published a deeper catalog of 
Centaurus galaxies including
a substantial dwarf population, using existing du Pont telescope photographic
plates covering 2.2 square degrees of the cluster in the
B-band.  They acertained cluster membership through morphological criteria.
\citet{sjf97} performed a follow-up spectroscopic study 
measuring redshifts for 115 galaxies to $B_{T} < 19.5$ including
32 new redshifts for dwarf galaxies.  
Their Centaurus Cluster Catalog (CCC) of likely cluster members was
used to select the spectra sample.
The majority of these galaxies were found
to be members with only a 12\% contamination from background galaxies.
However, it is unknown what fraction of cluster dwarfs they may have missed
when using morphological criteria to establish membership and define the
spectroscopic sample.

In this work, we obtain spectra for a random sample of galaxies in
the direction of the Centaurus cluster,
the only qualification being that they lie within prescribed surface brightness and
magnitude ranges.  This sample was selected from our photometric
observations of the central 0.83 degrees$^2$ of the Centaurus cluster,
the results of which are presented in Paper I \citep{KCMM1}.
This is part of our project to measure the LF faint-end slope in a large sample
of nearby clusters in order to establish whether environmental factors impact the
shape of the cluster LF and estimate the mass contribution to galaxy clusters by
the dwarf population.
The aim of this paper is to use kinematics to provide new information with 
which we can assess the
degree of background contamination as faint into the LF as possible.

In section \ref{obs} we discuss sample selection and detail the 
observations.  We provide a description of the data reduction
in section \ref{redc}. We present our radial
velocity measurements in section \ref{rvsec}, with a description of
how these are used to determine the galaxy LF of Centaurus in  
section \ref{lfsec}.  We conclude with a discussion and summary of our results
in section \ref{discsec}.
Throughout this work we assume H$_{\circ} = 70$ km s$^{-1}$ Mpc$^{-1}$.

\section{Observations\label{obs}}

Photometric observations of A3526 (Centaurus, z = 0.0114, $12^{h}48^{m}51.8^{s}$
$-41^{\circ}18^{\prime}21^{\prime\prime}$ 2000.0) were acquired with the
LCO 1m with a TEK 2K Camera having a field of view of 20$^{\prime}$.48 and
scale size $0^{\prime\prime}.6$/pixel.
The seeing ranged from 1$^{\prime\prime}$.18 to 1$^{\prime\prime}$.68
and 2 nights were photometric.
\citet{land92} standard stars were regularly observed
on these nights.
All data were taken in the V-band for maximum light transmission and
the cluster was observed in a mosaic pattern with each field exposed
for 4x15 minutes for a total exposure of 1 hour.
Eleven separate fields were observed to obtain a total coverage of 0.83 degrees$^2$.
The photometric sample is complete to M$_{V} = -12.4$ and to a surface brightness
of $\sim25$ mag arcsec$^{-2}$. The magnitude limit is due primarily to errors
in distinguishing between stars and galaxies at fainter magnitudes. 
Over 10,000 galaxies were detected, but only $\sim 2700$ to these limits.
The reduction process for these data is described in Paper I. 

A spectroscopic follow-up was made of $\sim$ 500 of the
brightest
galaxies in A3526 using the LCO 2.5m with a multi-fiber spectrograph and
2D-Frutti faint object detector. The fibers have a diameter of
3$^{\prime\prime}.5$
corresponding to a physical size of $\sim$ 800 pc at the distance of the
cluster.  A 600
line mm$^{-1}$ grating was used to obtain a spectral resolution of
$\sim$ 8.5 \AA \ based on FWHM measurements over
the region 3800-6500\AA.
Spectra were taken in 5 different setups with
each setup containing 128 fibers including 16 sky fibers.  Due to 
several bad fibers, there were about 106 fibers available in each
setup for galaxy spectra acquisition. The spectroscopic sample was chosen to include
a random selection of 500 of the brightest galaxies from our Centaurus region galaxy 
catalog. 
We did not wish to bias this sample by observing only those
galaxies we expected to be cluster members through, for example, morphological criteria.
The galaxies within a particular setup all had the same magnitude and surface
brightness range and total exposure times ranged from 2 to 10.6 hours depending 
on the magnitude limit in each setup.
Table \ref{table4} provides a summary of the spectroscopic observations.  An apparent
magnitude of 17 corresponds to M$_{V} \sim -16.5$ or to about the bright end
of the dwarf population.  The magnitudes listed in this table have not
been corrected for reddening which is substantial in the direction of
Centaurus.  Total magnitudes measured hereafter have been corrected to 
account for A$_{V} \sim 0.36$ magnitudes of extinction.

\section{Processing and Reduction Methods\label{redc}}

Photometric data were processed as described in Paper I.
The 1D spectra were reduced using the HYDRA package in IRAF. All images
were bias corrected and flattened.  Individual spectra were
extracted using a flat with high S/N to serve as a template
where each of the 128 spectra was fit with a 6th order
Legendre polynomial.  The spectra were then dispersion
corrected using arc lamp wavelength calibration spectra
taken both before and after each setup exposure. Finally, the spectra
were sky subtracted using data from the dedicated 16 sky fibers.

\begin{figure}[t]
\begin{centering}
\includegraphics[angle=0, totalheight=2.3in] {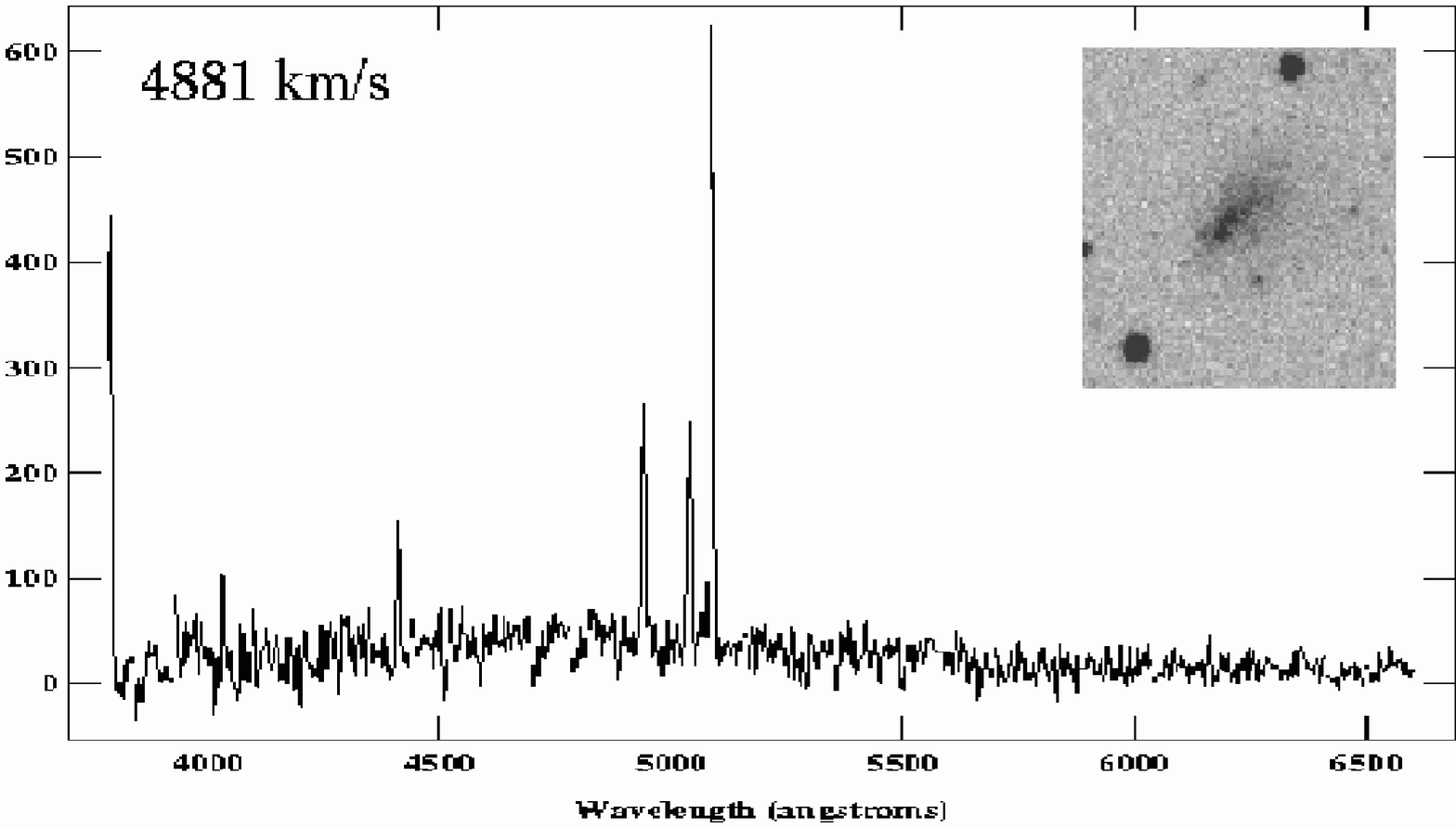}
\hfil
\includegraphics[angle=0, totalheight=2.3in]{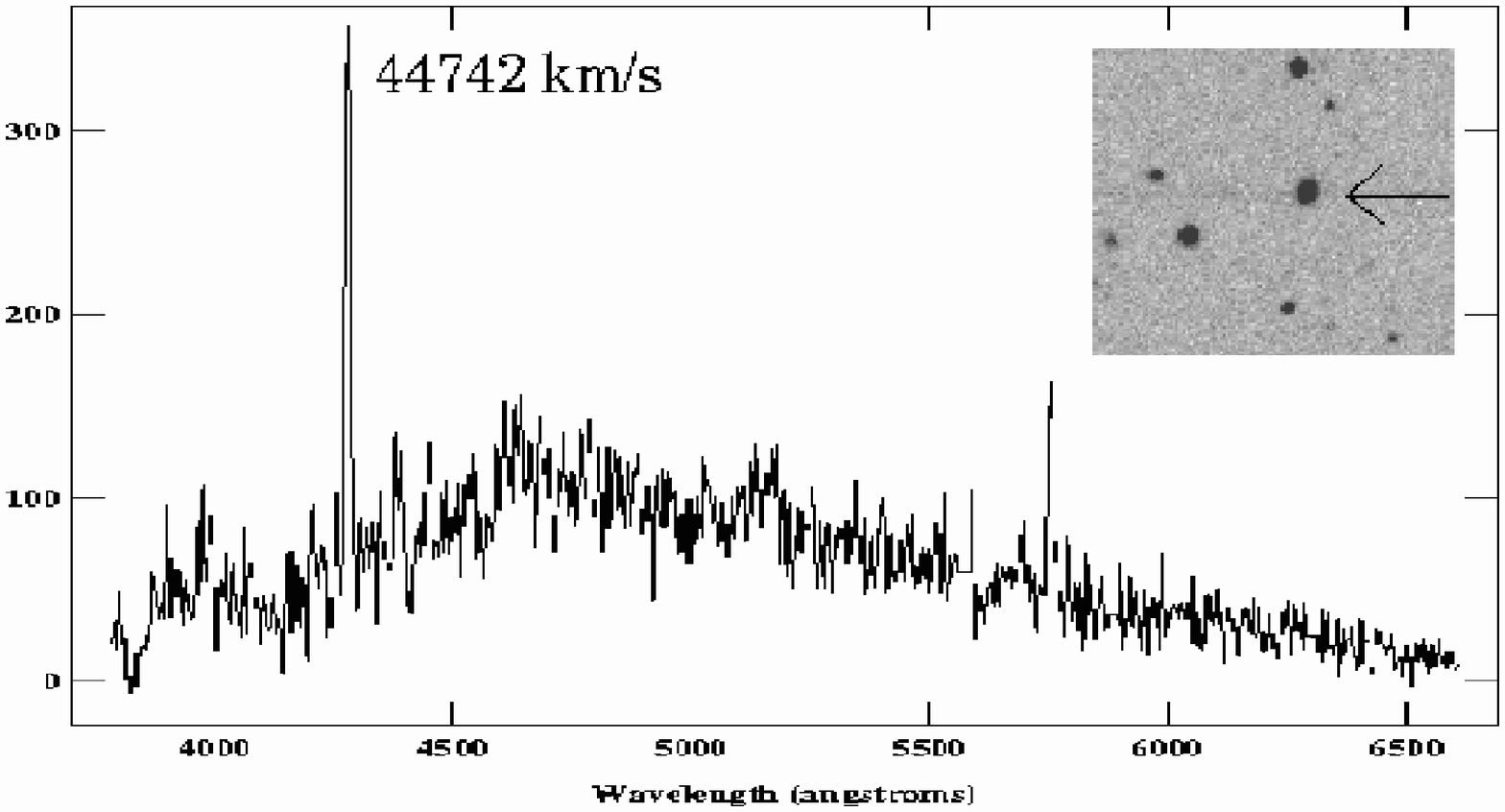}
\hfil
\includegraphics[angle=0, totalheight=2.3in]{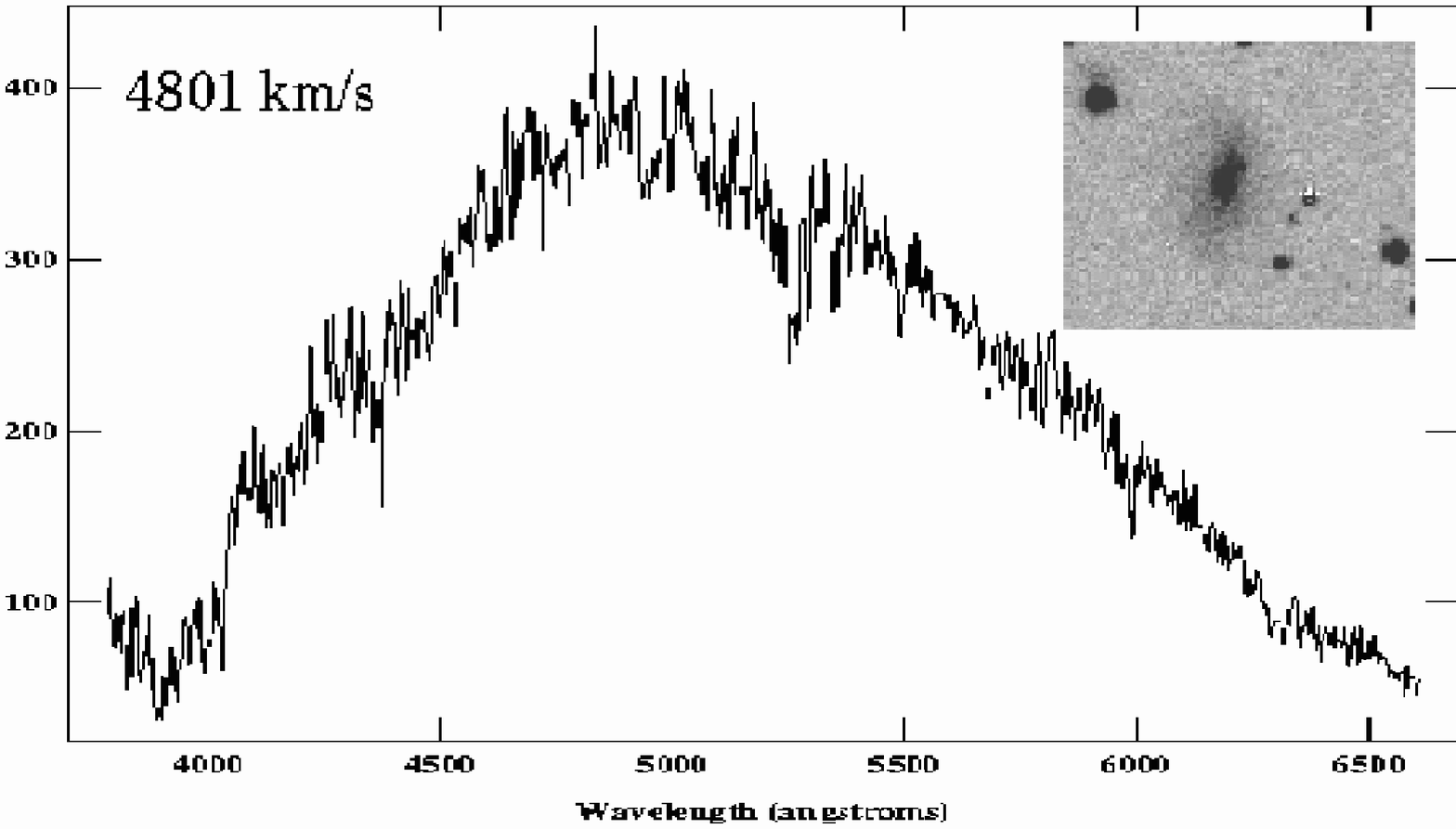}
\caption[Spectra for three galaxies]{Spectra for three galaxies.  The top
and bottom plots display spectra of cluster dwarfs, while in the middle plot 
the compact, star-like object proved to be at high
redshift.  In the top plot, we were able to extract a redshift for
the very LSB dwarf through its emission line spectrum.  For the dwarf
in the bottom
plot, we obtained a redshift from cross correlation measurements of
absorption lines due to the presence of a bright nucleus.
\label{spectplots}}
\end{centering}
\end{figure}

Radial velocities were determined via Fourier cross correlation with
a stellar template using the
task XCSAO in RVSAO 
\citep{km98} on absorption line spectra.
Velocity errors (derived from the 
\citet{td79} R
values and the width of the cross-correlation peak) 
typically ranged from 20-120 km/s. 
Many of the velocities obtained for low surface brightness
galaxies, however, were through emission line spectra.  These galaxy redshifts
were usually identified with O[II]$\lambda$3727, Ne[II]$\lambda$3869, and 
the triplet of
H$\beta$, O[III]$\lambda$4958.9, and O[III]$\lambda$5006.8 and velocities
were calculated
from an unweighted mean of all emission lines present.  We take the standard
deviation from these measurements as our error when at least three lines
are used.
Finally, a heliocentric correction was made for the
emission line redshifts.  A few examples of
the galaxy spectra are shown in Figure \ref{spectplots}.  

\section{Radial Velocities\label{rvsec}}

\begin{figure}[t]
\begin{centering}
\epsscale{0.75}
\includegraphics[angle=270, totalheight=3.50in]{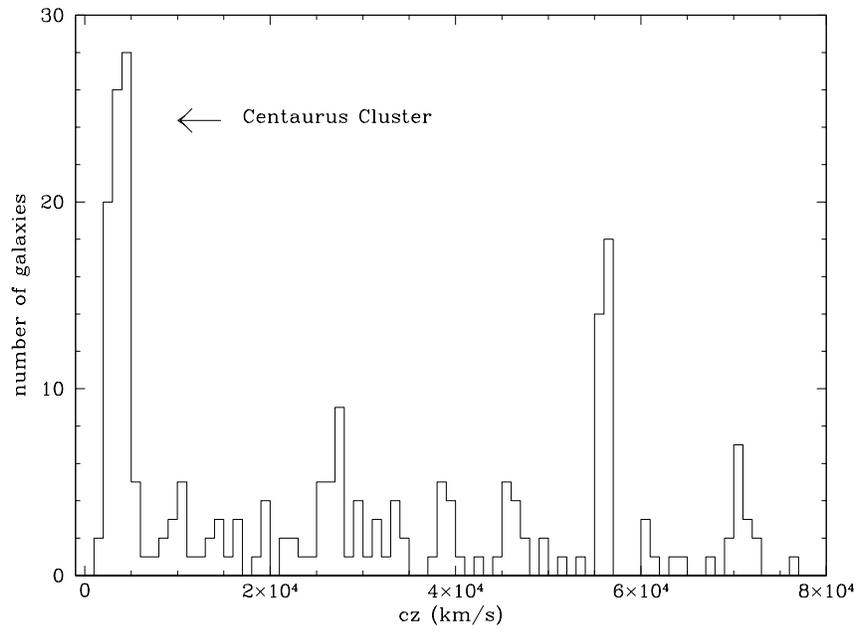}
\caption[Radial velocties of galaxies in Centaurus]{Histogram of the
radial velocities obtained from the spectroscopic data.  Several clusters or 
overdensities are evident behind Centaurus.
\label{vels}}
\end{centering}
\end{figure}

Each of the five setups contain at least 100 spectra and
the success rate for determining velocities drops from $\sim 65$\% at
$V \leq 17.0$ to $\sim20$\% at $V \geq 19.0$.  
Table \ref{table4} lists the number of cluster members and higher redshift galaxies
in each setup for which velocities were acquired. 
A total of 225 radial velocities were obtained 
to a limiting magnitude of $V < 19.5$ ($M_{V} < -13.9$) and with 
a limiting central surface brightness of $\mu_{\circ} = 23.0$ mag arcsec$^{-2}$.  
Radial velocities for cluster galaxies can be found in Table \ref{tablemem}
while those for background galaxies are provided in Table \ref{tablenon}.
A separate list of those cluster galaxies with new redshifts is given
in Table \ref{tablenew}.
A histogram of all the radial velocities obtained
is displayed in Figure \ref{vels}.
In the plot, the Centaurus cluster is obvious at low redshift (z = 0.011)
while higher redshift structures are apparent at cz = $3.8 \times 10^4$ 
and $5.6 \times 10^4$ km/s.  

Of the galaxies for which velocities
were obtained, we found 35\% to be member galaxies.  We
used a cutoff of $v_{max}$ = 5450 km/s for cluster members since this
is the $\sim3\sigma$ upper limit of the velocity distribution
for the higher velocity subcomponent in Centaurus (at 4500 km/s) \citep{sjf97}.
The $3\sigma$ lower limit is only 200 km/s and is therefore ignored.
New redshifts were obtained for 15 Centaurus dwarfs.  
Of the 78 member galaxies, magnitudes ranged from $11.8 < V < 18.5$
($ -21.6 < M_{V} < -14.9$) with central surface brightnesses brighter than 22.5 mag/arcsec$^{2}$.
We consider 25 galaxies with M$_{V} > -17.0$ to be dwarfs.
For some galaxies we have radial velocity measurements from both emission and
absorption lines. 
Comparing the results for 34 galaxies for which both measurements were
obtained, we find a
mean offset in v$_{abs} - $v$_{em}$ of 60.2 km/s, with $\sigma_{\Delta v} = 198.3$ and 
$\sigma_{\overline{\Delta v}} = 34.0$.

\begin{figure}[t]
\begin{centering}
\epsscale{0.75}
\plotone{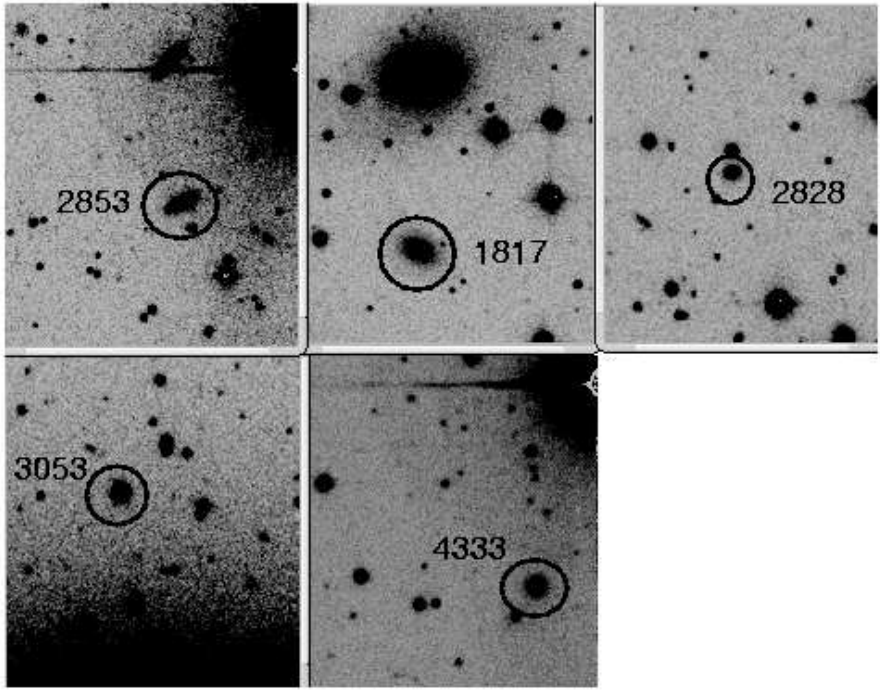}
\caption[Five galaxies for which new redshifts have been obtained]{
Five galaxies for which new redshifts have been obtained.  Radial
velocities (km/s) are printed next to the circled cluster dwarfs.
All of these
are likely Centaurus cluster members.  While
four of these (with the exception of the galaxy at 4333 km/s) are within
the survey limits of the Centaurus Cluster Catalog of \citet{jd97}, none
are listed as members in the catalog.
\label{compactg}}
\end{centering}
\end{figure}

Four cluster galaxies for which we obtained new redshifts
are not included in the Centaurus Cluster Catalog of \citet{jd97}.
These are all compact or compact
disk galaxies, which, through morphological means were not classified as cluster
members.  We display these galaxies 
in Figure \ref{compactg}, along with a fifth similar compact dwarf.   
From an all-object spectroscopic survey in Fornax,
\citet{djgp00} have found a new population of ultra compact
dwarf galaxies (UCDs) which are often overlooked in typical galaxy surveys.  This
underscores the importance of using spectra or some other means besides
morphology to establish cluster membership.

\begin{figure}[t]
\begin{centering}
\epsscale{0.88}
\plotone{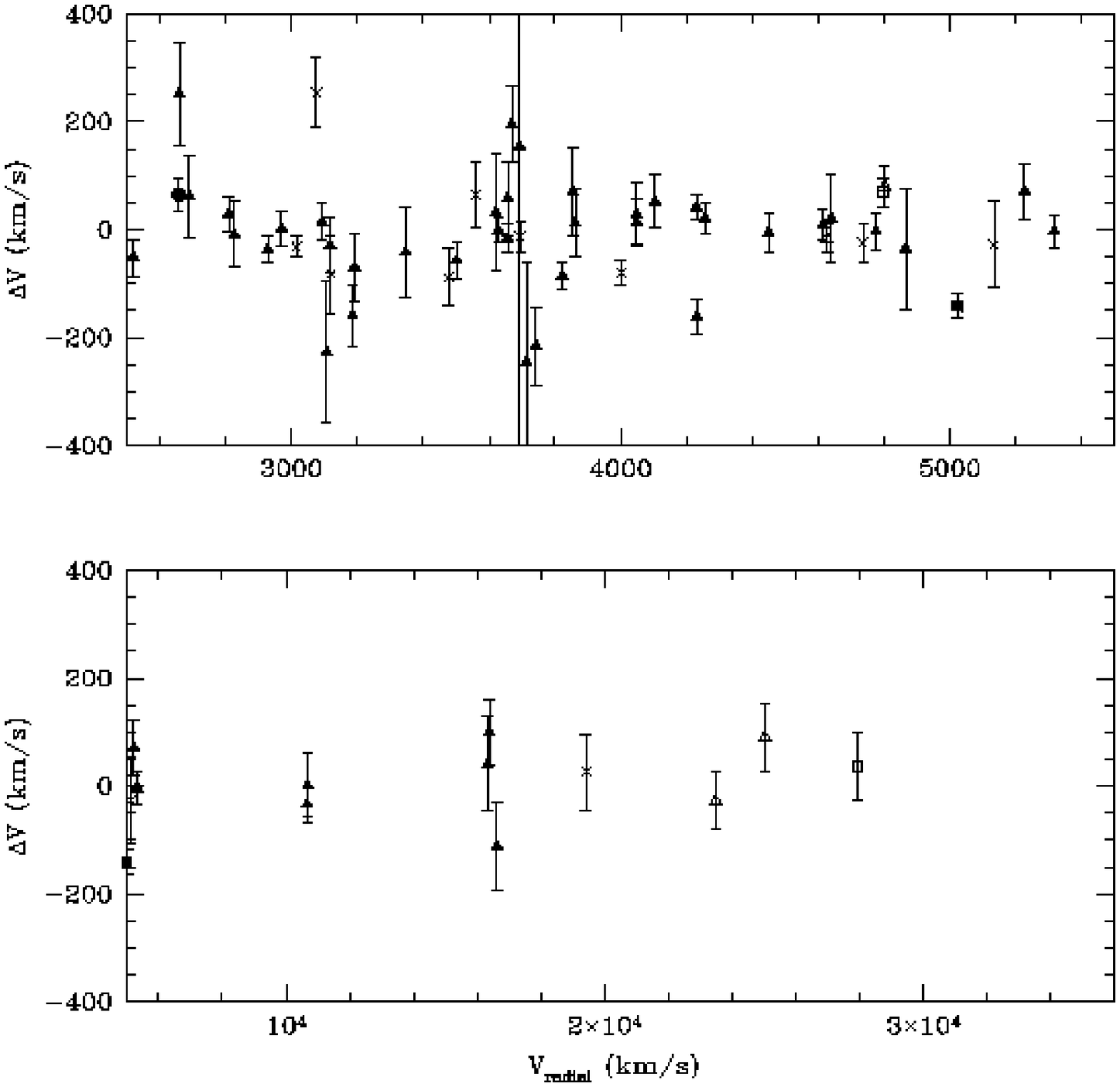}
\caption[Comparison of radial velocties with literature values]
{Comparison of radial velocties with literature values.  We plot
v$_{rad}$(literature) $-$ v$_{rad}$(this work) vs our radial velocity measurements.
Solid triangles correspond to \citet{sjf97}, solid squares to
\citet{dcl86}, solid circles to \citet{mf96}, crosses to \citet{s96},
open triangles to \citet{slhsd00}, and open squares to \citet{lv89}.
\label{velcomp}}
\end{centering}
\end{figure}

In Figure \ref{velcomp}, we compare our velocities to those in the
literature\footnote{This research has made use of the NASA/IPAC Extragalactic
Database (NED) which is operated by the Jet Propulsion Laboratory, California
Institute of Technology, under contract with the National Aeronautics and
Space Administration.}.  The majority of velocities for cluster dwarf galaxies come
from \citet{sjf97}.  
We find good agreement with the literature
values with a mean difference between our values and the literature of
$-4.3$ km/s with $\sigma = 97.8$ and $\sigma_{\overline{\Delta}} = 11.7$.  
A comparison of our data to just those
of \citet{sjf97} yields a mean difference of $-6.2$ km/s with $\sigma = 105.6$ and
$\sigma_{\overline{\Delta}} = 15.1$.

\begin{figure}[t]
\begin{centering}
\epsscale{0.75}
\plotone{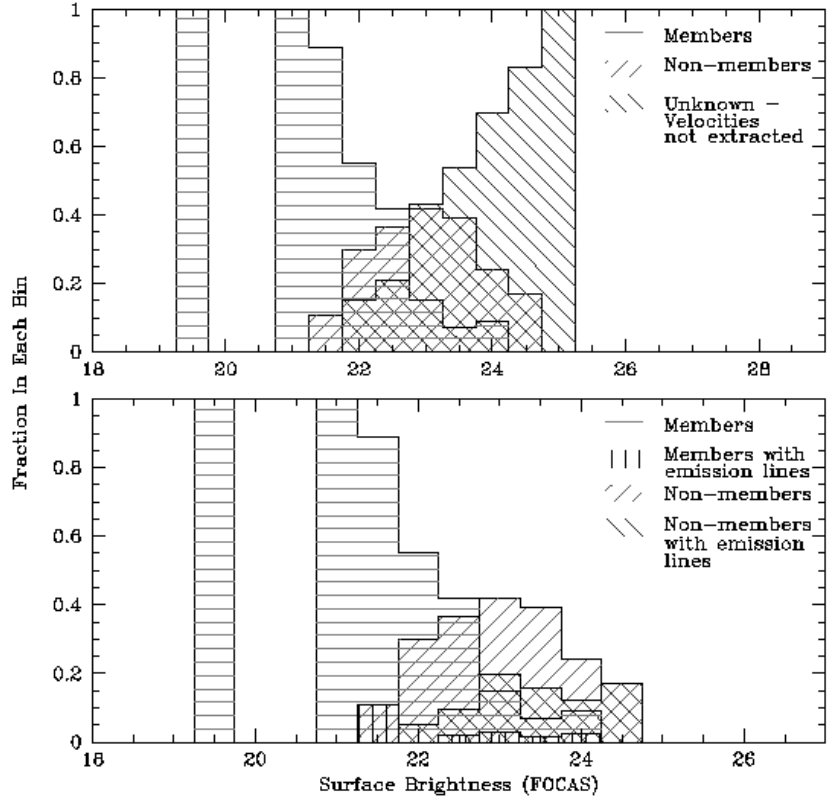}
\caption[Centaurus cluster membership from redshifts vs. surface brightness]
{Centaurus cluster membership determined from spectra.  We display the fraction
in each surface brightness bin determined to be members or non-members.
The surface brightness in this plot is the total surface brightness
over the isophotal detection area.  In the lower plot we further
split detections into spectra with and without emission
lines.
\label{srecovery}}
\end{centering}
\end{figure}

In Figure \ref{srecovery} we display the recovery of redshifts as a
function of mean isophotal surface brightness.  We find that all the highest
surface brightness galaxies are cluster members after which the cluster
fraction exhibits a sharp drop-off.  This is in large part due to the
difficulty in obtaining spectra for low surface brightness objects.  The
fraction of galaxies for which we obtained spectra but were unable to
extract redshifts begins to rise at a surface brightness of 22.0 mag arcsec$^{-2}$, and
by 25.0 mag arcsec$^{-2}$ we are unable to measure any redshifts.  We note that we are able to
measure redshifts for more distant galaxies a half magnitude fainter
than the cluster members.  This may be due to the
presence of emission lines in the non-cluster galaxies.  Emission
lines can be used to
determine redshifts even when the galaxy continuum is too low to measure
and the spectrum too noisy to use absorption lines.
Cluster galaxies tend to have less gas and lower star formation rates, particularly
those galaxies near the cluster core, and are
therefore less likely to have emission line spectra \citep{ghc82,bkmmhpf97}.
Thus, we also distinguish in Figure \ref{srecovery}
between the fractions recovered with and without emission lines.
We find that in the faintest surface brightness bin where
we obtain redshifts for distant galaxies only, all velocities are obtained through
emission lines.
                                                                                      
\section{Luminosity Function\label{lfsec}}

\begin{figure}
\begin{centering}
\epsscale{1.0}
\includegraphics[angle=270, totalheight=3.50in]{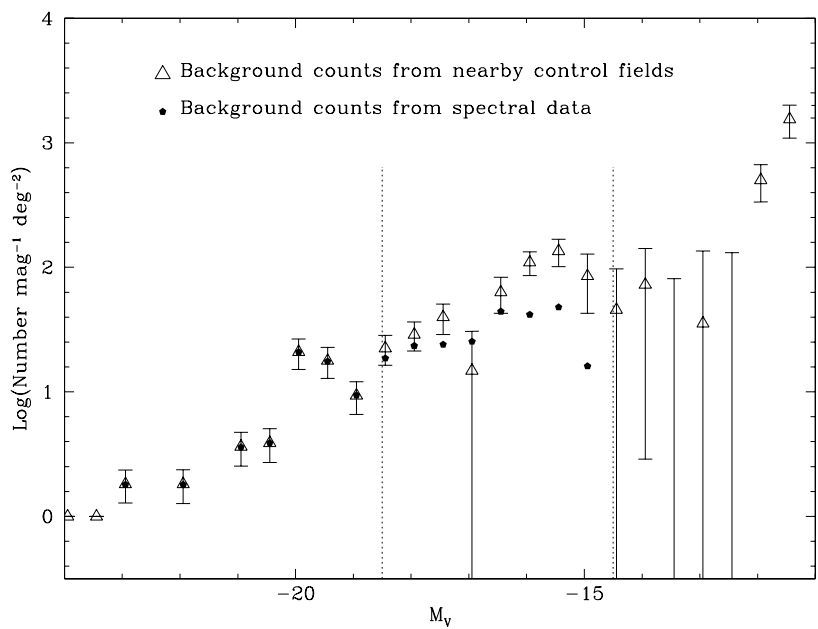}
\caption[Centaurus cluster LF from spectroscopic results]
{Centaurus cluster LF from spectroscopic results.  Open triangles
show cluster member counts after subtraction of local control field counts
while solid symbols represent counts
determined from redshifts.  No correction is made for the spectra which
did not yield redshift measurements.  For clarity,
error bars are only supplied for the former method.  Neither spectra
nor control field counts find background galaxies with $V < 15$
(all galaxies brighter than M$_{V} = -18.4$ are members),
and member velocities were only obtained to M$_{V} \sim -15.0$.
\label{spectlf}}
\end{centering}
\end{figure}

We can use our kinematic results to help improve on our estimate of background
contamination in our photometric determination of the Centaurus LF.  
We first determined
the fraction of members to the total number of galaxies for which
redshifts were obtained in each half magnitude bin.
Using these fractions, along with the binned
number of galaxies per square degree detected in the cluster fields, we 
calculated the expected
number of members in each half magnitude bin. In Figure \ref{spectlf} we
show the LF constructed using the redshift determined member fractions along with the
LF derived from the background subtraction method
using nearby control fields described in Paper III \citep{KCMM3}.  At magnitudes brighter
than M$_{V} < -18.4$, neither the control fields nor the spectra turned up any background
galaxies.  These points are therefore identical.  Member counts from 
redshifts only extend to M$_{V} = -15.0$.
The cluster galaxy counts obtained in this way
fall below counts calculated statistically.
This is likely because our calculated fraction of member dwarf galaxies
to background galaxies is too small due to either
the lower surface brightness of the dwarfs,
or because the background galaxies more often have emission lines and
are preferentially recovered.

\begin{figure}
\begin{centering}
\epsscale{1.0}
\includegraphics[angle=270, totalheight=3.50in]{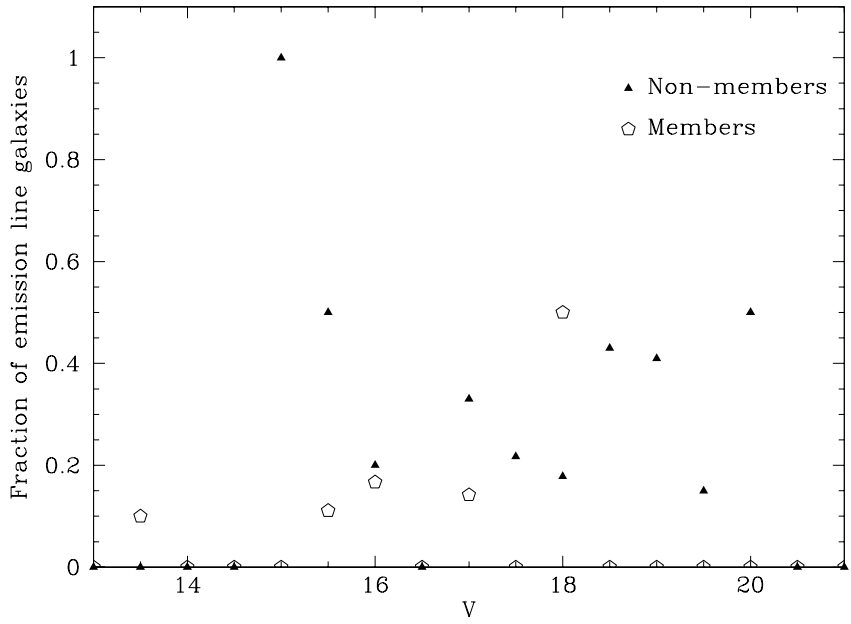}
\caption[Fraction of galaxies with v$_{rad}$ measured from emission line
spectra]
{Fraction of galaxies with v$_{rad}$ measured from emission line
spectra in each half magnitude bin.
Solid triangles represent the fraction of background galaxies
with radial velocities determined from emission line spectra while
open pentagons represent this fraction for cluster members.
\label{emiss}}
\end{centering}
\end{figure}

In Figure \ref{emiss}, we plot the fraction of member and non-member
galaxies
in each magnitude bin with radial velocities that were measured with 
emission line spectra.  In the faintest magnitude bins, where only 
high redshift galaxies are recovered, 40 - 50\% of these
had emission lines.  This, of course, means than at least half were recovered
through absorption line spectra and it is thereore possible to measure
redshifts at these magnitudes without the presence of emission lines.
It also shows a lack of emission line spectra in dwarf galaxies at these
magnitudes.  

\begin{figure}
\begin{centering}
\epsscale{1.0}
\includegraphics[angle=0, totalheight=3.50in]{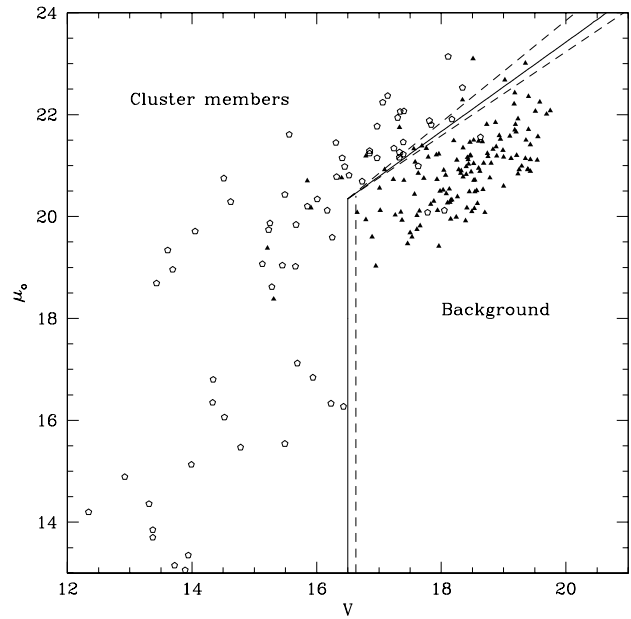}
\caption[Central surface brightness vs. total magnitude]
{Central surface brightness vs. total magnitude for galaxies with
measured redshifts.  Open symbols
represent cluster members while solid symbols indicate background
galaxies.  Lines are drawn which roughly separate these two
populations.  The solid lines define the surface brightness-magnitude
criteria we use to determine membership.  They are intentionally drawn
to include as many member galaxies as possible. Dashed lines provide
other possible delineations consistent with the data.  
\label{msbrec}}
\end{centering}
\end{figure}

It is likely that
we are unable to extract velocities from the spectra for a large 
population of low surface brightness non-emission
line cluster members.  We plot the central surface brightness vs. 
total magnitude for all galaxies for which we obtained spectra in
Figure \ref{msbrec}.  The central surface brightnesses were determined
from surface brightness profile fitting as described in Paper I
and are uncorrected for seeing
effects.  We can define a region in this 
$\mu_{\circ} - V$ space which encloses most member galaxies.  It is
apparent from this plot that most dwarf galaxies have
much lower surface brightnesses than background galaxies at a given
magnitude.  This could account for the fewer cluster galaxies recovered
by the spectra than would be expected from the LF counts determined using
statistical background correction methods. 

\begin{figure}
\begin{centering}
\epsscale{1.0}
\includegraphics[angle=0, totalheight=3.50in]{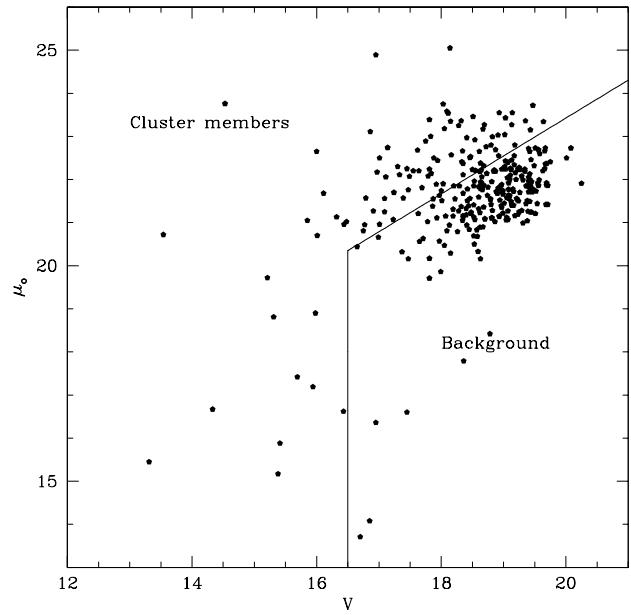}
\caption[Central surface brightness vs. total magnitude for galaxies
for which redshifts were not obtained]{$\mu_{\circ}$ vs. $V$
for 306 galaxies for which redshifts were not obtained from the spectra.
Note, the scale has been changed to include even lower surface brightness
galaxies.
\label{msbunk}}
\end{centering}
\end{figure}

\begin{figure}
\begin{centering}
\epsscale{1.0}
\includegraphics[angle=270, totalheight=3.50in]{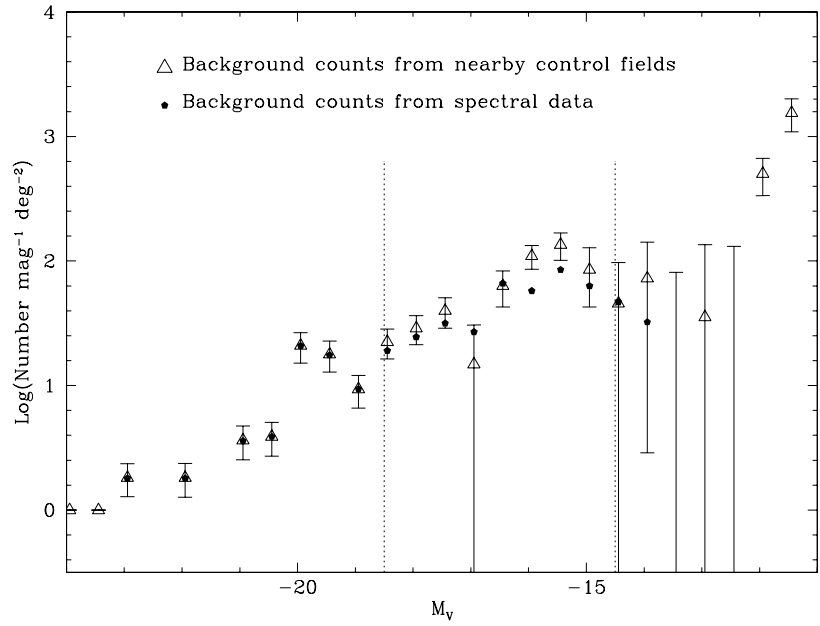}
\caption[LF with correction for galaxies
for which redshifts were not obtained]{LF with correction for galaxies
for which redshifts were not obtained.
Again, the open triangles
refer to member counts corrected by statistical background subtraction while
the solid symbols are counts determined from spectra, now taking into
account the 306 galaxies which did not produce measurable spectra.
\label{lfunk}}
\end{centering}
\end{figure}

We therefore use these surface brightness regions to make a correction 
for the large number
of galaxies for which we were unable to extract redshifts.  
Central surface brightnesses and magnitudes for galaxies for which 
we obtained spectra but were unable to measure redshifts are 
plotted in Figure \ref{msbunk}.
For 
simplicity, we assume that all galaxies lying within the region 
denoted as `Background' are background galaxies and the
rest are cluster members.  While there will be some cross-contamination,
it does not appear to be severe and, if anything, will overestimate the
number of cluster members.  After including these counts as
members or background galaxies, we again determine the fraction of 
cluster members
in each magnitude bin and use these to calculate the total number of cluster
members from the total observed counts in each magnitude bin.  The
LF is shown in Figure \ref{lfunk}.

\begin{figure}
\begin{centering}
\epsscale{1.0}
\includegraphics[angle=0, totalheight=3.50in]{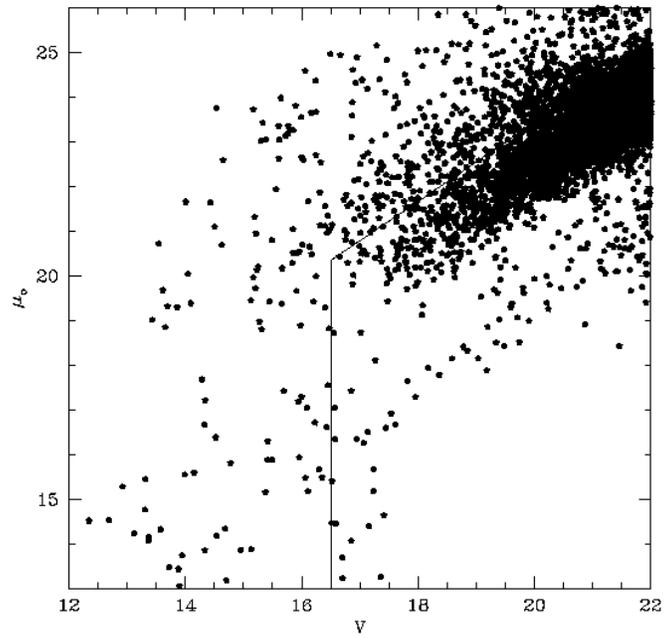}
\caption[$\mu_{\circ} - V$ for all galaxies]
{$\mu_{\circ} \ vs. \ V$ for all galaxies within
the central 0.83 degrees$^{2}$ of the Centaurus cluster down to
$\mu_{\circ} = 26$ mag / arcsec$^2$.  The lines separating the
region of background from cluster galaxies are also displayed.
\label{allgal}}
\end{centering}
\end{figure}

We find that counts are now only slightly lower than those calculated statistically.
A $\mu_{\circ} - V$ 
plot of all galaxies found in the Centaurus fields (Figure \ref{allgal})
reveals that there is
still a population of galaxies with even lower surface brightness beyond
those for which we tried to obtain spectra.  
We should therefore still be underestimating the dwarf population.

\begin{figure}
\begin{centering}
\epsscale{1.0}
\includegraphics[angle=0, totalheight=4.50in]{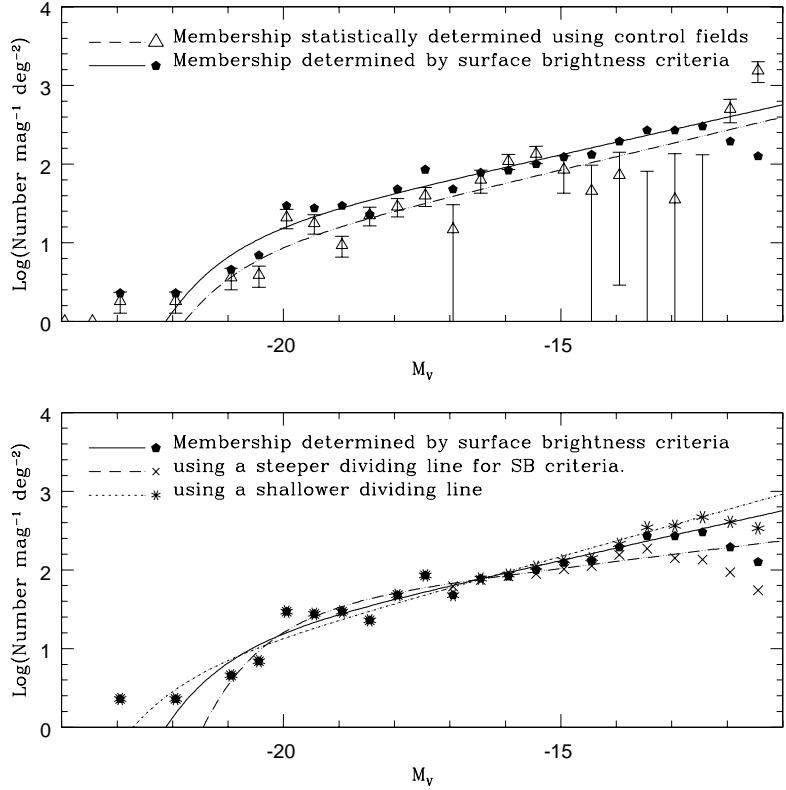}
\caption[Centaurus cluster LF]
{Centaurus cluster LF.  Total counts are corrected for
background contamination either with
a statistical correction (open points), or through surface brightness-magnitude
criteria (solid and skeletal points).   In the upper plot, the dashed line is the best 
Schechter function
fit ($\alpha = -1.42\pm0.2$) for the statistical LF, while the solid line is
the best fit for the surface brightness constructed LF, with $\alpha=-1.40$.
In the lower plot,
we test the effect of steepening and lowering the faint-end division line
for this latter case and find the data are consistent with slopes as
steep as $-1.50$ and as shallow as $-1.22$. The fits only apply to
$V < 21$ (M$_{V} < -12.4$).
\label{finLF}}
\end{centering}
\end{figure}

As one final estimate of the LF, we use the $\mu_{\circ} - V$ regions
to define the membership status for each galaxy from our complete catalog
of Centaurus galaxies (see Paper I).  
In order to do so, we must extrapolate the surface brightness selection line
by about 2.5 magnitudes to reach our photometric survey limits.  At fainter 
magnitudes and surface brightnesses, errors are larger so we expect the
contamination to increase as larger errors cause points to migrate relative 
to the selection line.
In Figure \ref{finLF}, we display the luminosity function
for galaxies determined in this manner to be cluster members. 
Counts are now slightly higher than those calculated through statistical
means, but only around the $1\sigma$ level.  
Differences are expected because the two LFs are constructed
in different ways. The original LF counts are weighted averages which include
corrections for completeness and misclassification whereas here we use the 
complete catalog to directly determine membership.
It is also likely that we
are overestimating the member fraction as the contamination from background
galaxies in the low surface brightness region appears to be slightly greater
than of cluster galaxies in the high surface brightness region (see
Figure \ref{msbrec}).
A best fit using a Schechter function \citep{s76} for
the statistical LF yields a faint-end slope, $\alpha$, of  $-1.42$ and M$_{V}^{*}$ of $-21.25$ 
while a power law fit to just the
faint-end finds a best fit slope of $-1.40$.  For a complete 
description of the measurement of this LF, see Paper III. 
The slope of the LF using surface brightness criteria does
not differ significantly from the statistical LF.  While the
normalization is slightly higher, this LF is
best fit with a Schechter function having $\alpha = -1.40$ and M$_{*} = -21.22$.
The statistically determined LF exhibits a drop-off
in counts in the magnitude range $-15 <$ M$_{V} < -12$ followed by a steepening at the
faintest magnitudes which is not mimicked by the LF established with surface 
brightness criteria.  
Rather, this LF maintains a constant slope until a turnover due to incompleteness at
M$_{V} = -12$. 

This surface brightness criteria LF we have constructed is 
presumed to be an overestimate of member galaxies since we have drawn the
division lines to include as many members as possible.  
However, it is possible, given the few data points in Figure \ref{msbrec}
which we use to establish the membership criteria,
to use other delineations which are still consistent with the data.
To determine the systematic uncertainty in the faint-end slope using this method, we
redraw the division lines in Figure \ref{msbrec} such that they
minimize and maximize
the number of galaxies labelled as background while remaining consistent
with the spectroscopy results.  When measuring the lower limit of the faint-end
slope, we
also push the bright-end division line to fainter magnitudes, 
consistent with the data.  This produces a lower (shallower) value for the 
faint-end slope of $\alpha=-1.22$.  We find an upper limit to the slope
of $-1.50$.  Random errors such as magnitude or surface brightness 
measurement errors will increase this uncertainty.

\section{Discussion\label{discsec}}

We have used spectroscopic redshifts of 225 galaxies to assist in 
constructing the Centaurus cluster galaxy LF and compare to our photometric LF 
obtained 
with a statistical background correction using nearby control fields.
We find counts and slope lower when strictly using the spectroscopic results 
than from the statistically corrected photometric sample.  
However, it is clear that distant late
type emission line and higher surface brightness galaxies are easier to obtain 
spectra for than low surface brightness (LSB) cluster dwarfs at the same faint magnitudes.  
When we make use of the separation between distant and member galaxies on a magnitude -
surface brightness plane, we find a similar faint-end slope compared  
to the control field subtracted LFs ($-1.40$ vs $-1.42\pm0.2$) although 
the faint-end slope for the $\mu_{o} - V$ division case could
realistically range from $-1.22$ to $-1.50$. Errors using
either method are similar and both may suffer from similar systematics
(see below).

First we explore the cause of the $\mu_{o} - V$ division between background 
and cluster galaxies.  Background galaxies appear to primarily lie 
in a clump within a small range of surface brightness and magnitude
(between $19 < \mu_o < 22$ and between $16.5 < V < 20$).  
If we take a bright elliptical galaxy in Centaurus with a $V$ magnitude of 
13.4 (M$_{V} = -20$), 
central surface brightness of 13.9 mag/arcsec$^2$, and half-light radius, 
R$_e$, of $12^{\prime\prime}.2$ and shift it to a redshift of 0.1 (about the distance
of the first large background structure), we find that Re reduces to
$1^{\prime\prime}.5$, which is approximately the average seeing for our
survey. According to \citet{tacg01}, even in the case of our best seeing 
image, $1^{\prime\prime}.1$,
Sersic profile fitting should recover R$_{e(rec)} \sim 3.5$ R$_{e(true)}$ and
I$_{o(rec)} \sim 0.0015$ I$_{o(true)}$ when this seeing is
not taken into account.  The central surface brightness for this
galaxy would accordingly be inflated to 20.9 mag/arcsec$^2$ with V $= 18.3$, within
the range of the background clump.
Thus, the separation between background and cluster galaxies in 
our surface brightness $-$ magnitude plots is an artifact of seeing where
the cluster members are primarily those galaxies which are 
resolved.  Using this surface brightness $-$ magnitude criteria for membership
we do not expect to exclude many dwarf galaxies unless there is a large
population of very compact dwarfs, nor do we expect a large population
of giant resolved background galaxies to contaminate the cluster member region.

While the method of using redshifts to establish cluster membership should be more accurate
than the statistical method for estimating the background contribution,
we find that it does have shortcomings.  First, there is a selection bias in that galaxies
for which spectra are taken must be chosen ahead of time.  Since even
multi-fiber spectrographs have a limit to the number of spectra which can be
obtained for each exposure and exposures for faint, low surface brightness
galaxies must be
long, often there is a selection bias for choosing galaxies which are
likely to be cluster members.  Although we randomly chose our sample based
entirely on magnitude and surface brightness ranges, \citet{sjf97},
for example, select their
spectroscopic sample from their Centaurus Cluster Catalog which
contains galaxies likely to be Centaurus cluster members based on
morphological criteria. As we have found, some compact cluster
members may be missed in this way.  
Furthermore, low surface brightness and highly compact
galaxies must first be cataloged in photometric
studies and it is possible that an entire population of low surface brightness
dwarf galaxies may be missed.  

Second, and more importantly, spectra can only be
obtained for brighter, high surface brightness galaxies or for galaxies with
emission lines (unless much larger telescopes are used). 
Spectroscopic studies finding fewer
galaxies than might be expected at fainter magnitudes are likely simply
unable to obtain measurable spectra
for large numbers of dEs with low surface brightnesses and low star
formation rates.  Our spectra yield redshifts at the faintest surface brightnesses
for emission line background galaxies only.  We find that we are unable
to extract velocities from spectra for a large fraction of low surface
brightness galaxies in each magnitude bin which are likely, based on our
spectra results, to be preferentially member galaxies.  Thus, any
spectroscopic study of nearby cluster galaxies will be systematically 
biased toward recovering redshifts for background galaxies.

A further important consideration is the depth of the study.
\citet{dep03} measure a faint-end slope for cluster
galaxies of $-1.28\pm{0.03}$ from the 2dF galaxy redshift survey down to
a limiting magnitude of M$_{b_{J}} < -15$. While they do not
find a steep slope, they also have a fairly bright limiting magnitude,
similar to this work.
Many photometric studies which find steep slopes probe to much fainter
limiting magnitudes (eg. \citet{t97,tcoma98,ppsj98,dphm95}).  If a steeper upturn does exist at faint magnitudes,
it would be difficult to observe with spectroscopic studies which
do not reach these fainter magnitudes.

While there are limitations to using spectra to establish membership,
constructing the LF through statistical means faces great uncertainties due to
field-to-field variance.
This method assumes that background galaxy counts are similar in
both cluster and control fields.  However, it is known that
counts can vary from field to field by as much as $19 - 50$\% on scales of
$0^{\circ}.4-0^{\circ}.5$ degrees (\citet{hjddgo00} and
references therein).  In addition, clusters which are originally discovered by
eye may be biased towards regions of greater density contrast with their surroundings.
\citet{vmml03} argue that Abell clusters may have been picked out due to 
projection effects and suffer from severe background contamination.
While the spectra results presented here for Centaurus, Abell 3526,   
corroborate the LF slope measured statistically,
this is an X-ray bright cluster. It may be worth using a similar
technique on other nearby, non X-ray clusters to determine whether 
statistically corrected LF faint-end slopes of these clusters are systematically
inflated.

The LF has been studied in numerous nearby clusters using various means to distinguish
between members and nonmembers. 
These studies have turned up a wide range of values
for the faint-end slope spanning $-2.3 < \alpha < -1.3$ (see e.g.
\citet{t97, ttv01, dphm95, vnml97} and \citet{bntuw95}).  
Whether this implies intrinsic differences exist between galaxy formation and/or 
evolutionary histories in these clusters or whether the differences are simply due to
the different methods used to construct the galaxy LF and in dealing with
the selection effects inherent in detecting low surface brightness
dwarf galaxies remains an unanswered question.  
We find no evidence of a very steep slope from either method employed here.
While it is possible that we could be missing a large population of LSB
dwarf galaxies, we do not believe this to be the case as we discuss in
greater detail in paper III.

To address the question of whether the cluster LF faint-end varies or has a universal
slope, it is critical to establish the faint-end slope in a number of clusters.
Correcting for background contamination statistically using
nearby control fields where counts are expected to vary from the true 
cluster background is fraught with large uncertainties.  
To accurately construct a cluster LF, membership must be determined
to faint limiting magnitudes directly.
For higher redshift clusters, it may be possible to use photometric redshifts
to establish membership (see ie. \citet{tmrldns04}).  
For the nearest clusters, namely Virgo, Fornax, and
Centaurus, some researchers have chosen to establish membership based on 
morphology.  We find that some compact galaxies may be missed in this way, although
only at around the 3-6\% level.  While UCDs are known to exist in clusters \citep{djgp00}, 
these galaxies
have magnitudes of around M$_V = -12.0$, typically at or below the magnitude limit
of most surveys. \citet{sjf97} found only a 12\% contamination rate of background
galaxies in their morphologically determined cluster sample using follow-up spectroscopy,
so membership determined in this manner is likely to be more reliable than
through statistical approximation.
For other nearby clusters, other methods for measuring distances to
the dwarf galaxies and establishing cluster membership need to be
devised, or spectra must be 
acquired to fainter limiting magnitudes and much lower surface brightnesses
with more sensitive instruments and larger telescopes
to firmly establish the faint-end slope in at least a few clusters.
While it is very difficult to obtain spectra to the necessary surface
brightnesses,
we demonstrate that it may be possible to use spectra in conjunction with
other properties to obtain better estimates of the nearby galaxy cluster LF.


\acknowledgments

We thank the anonymous referee for very helpful suggestions which
have improved the analysis and presentation of this paper.
KC would like to thank Joe Mohr for providing the stellar template for
extracting radial velocities from absorption line spectra.



\clearpage


\input{tab1.tex}

\clearpage

\input{tab2.tex}

\clearpage
                                                                                  
\input{tab3.tex}

\clearpage
                                                                                  
\input{tab4.tex}


\clearpage

%
%

\clearpage

\clearpage

%

\clearpage

\clearpage

%
                                                                                  
\clearpage

\clearpage

\clearpage

\clearpage

\clearpage

\clearpage

\clearpage

\end{document}

%% file: tab1.tex
\begin{deluxetable}{llllll}
\tabletypesize{\normalsize}
\tablewidth{0pt}
\tablecaption{Spectroscopic Observations\label{table4}}
\tablehead{
\colhead{Setup} &
\colhead{Exposure Time} &
\colhead{V range}      &
\colhead{$\mu_{o}$} & \colhead{\# Cluster} &
\colhead{\# Higher} \\
\colhead{ } &
\colhead{(hr)} &
\colhead{ } &
\colhead{(mag/arcsec$^{2}$)} &
\colhead{members} &
\colhead{redshift} \\
}

\startdata

100 & 2 & 12-17 & 20-23 & 57 & 8 \\
300 & 7 & 17-18 & 21-23 & 15 & 49 \\
500 & 6 & 18-19 & 21-24 & 2 & 39 \\
700 & 9 & 18-19 & 22-23 & 3 & 31 \\
900 & 10.6 & 19+ & 23-24 & 2 & 20 \\ \hline
Totals &  &  &  & 78 & 147 \\

\enddata

\end{deluxetable}

%% file: tab2.tex


\begin{deluxetable}{llllllllll}
\tabletypesize{\scriptsize}
\tablewidth{0pt}
\tablecaption{Cluster Members\label{tablemem}}
\tablehead{
\colhead{$\alpha$} &
\colhead{$\delta$}      & 
\colhead{$V$} & 
\colhead{$v_{rad}$} &
\colhead{Error} &
\colhead{$v_{rad}$} & 
\colhead{Error} &
\colhead{$v_{rad}$} &
\colhead{Error} &
\colhead{reference\tablenotemark{b}} \\
\colhead{2000.0} &
\colhead{} &
\colhead{total\tablenotemark{a}} &
\colhead{Absorption} &
\colhead{} &
\colhead{Emission} &
\colhead{} &
\colhead{Literature} &
\colhead{} &
\colhead{} 
}

\startdata

12 47 30.15 & -41 39 02.60 &  16.23 &  3743 &   73 &      &      &  3527 &   59 &  1 \\
12 47 33.42 & -41 07 54.70 &  15.94 &  3647 &  130 & 3740 &  160 &  3848 &   50 &  1 \\
12 47 45.05 & -41 09 21.30 &  16.52 &  2689 &   76 &      &      &  2750 &   43 &  1 \\
12 47 49.22 & -40 59 54.90 &  15.85 &  3186 &   58 &      &      &  3026 &   29 &  1 \\
12 47 52.20 & -41 20 13.60 &  18.05 &  2828 &   88 &      &      &       &      &    \\
12 47 54.18 & -41 03 44.90 &  15.56 &  4464 &   60 &      &      &       &      &    \\
12 47 55.84 & -40 54 16.50 &  17.33 &  1817 &   74 &      &      &       &      &    \\
12 47 56.01 & -40 55 14.00 &  14.52 &  3862 &   63 &      &      &  3874 &   70 &  1 \\
12 47 58.97 & -41 11 21.30 &  15.25 &  3117 &   50 &      &      &  3088 &      &  1 \\
12 47 59.57 & -41 13 11.40 &  14.78 &  4048 &   43 &      &      &  4063 &   19 &  1 \\
12 48 02.02 & -41 18 19.10 &  16.73 &  2473 &  101 &      &      &  2329 &   33 &  1 \\
12 48 07.29 & -41 37 46.30 &  18.34 &  4188 &  106 &      &      &       &      &    \\
12 48 15.45 & -41 42 57.20 &  16.97 &  3855 &   83 &      &      &  3925 &   45 &  1 \\
12 48 15.82 & -41 18 57.70 &  17.84 &       &      & 4256 &   28 &  4278 &   58 &  1 \\
12 48 16.68 & -41 26 10.60 &  17.34 &  2118 &   90 &      &      &       &      &    \\
12 48 22.75 & -41 07 23.50 &  13.90 &  3657 &   27 &      &      &  3641 &   49 &  4 \\
12 48 30.72 & -41 01 18.10 &  14.95 &  4231 &   31 &      &      &  4069 &      &  1 \\
12 48 31.00 & -41 18 23.60 &  15.13 &  2971 &   32 &      &      &  2973 &   44 &  1 \\
12 48 36.06 & -41 26 23.40 &  16.85 &  3346 &   83 &      &      &  3304 &   60 &  1 \\
12 48 39.69 & -41 16 05.60 &  16.32 &  2659 &   95 &      &      &  2910 &   67 &  1 \\
12 48 43.43 & -41 38 37.40 &  14.05 &  2240 &   30 &      &      &  2193 &   19 &  1 \\
12 48 48.56 & -41 20 53.50 &  17.78 &  3053 &   40 &      &      &       &      &    \\
12 49 02.00 & -41 15 33.30 &  17.39 &  2075 &   70 &      &      &  1958 &   71 &  1 \\
12 49 03.16 & -41 23 29.50 &  15.49 &  5227 &   52 &      &      &  5298 &   28 &  1 \\
12 49 04.12 & -41 20 19.60 &  14.51 &  3714 &  185 &      &      &  3469 &   34 &  3 \\
12 49 09.63 & -41 11 33.90 &  16.43 &  3192 &   64 &      &      &  3122 &   29 &  1 \\
12 49 12.04 & -41 32 40.80 &  14.33 &  2809 &   33 &      &      &  2838 &   20 &  6 \\
12 49 18.59 & -41 20 07.80 &  15.28 &  3093 &   33 &      &      &  3108 &   19 &  1 \\
12 49 22.68 & -41 15 18.60 &  18.17 &  3620 &  108 &      &      &  3652 &   61 &  1 \\
12 49 25.43 & -41 25 45.80 &  15.67 &  4048 &   58 &      &      &  4076 &   29 &  1 \\
12 49 26.21 & -41 29 20.80 &  13.37 &  4230 &   22 &      &      &  4271 &   21 &  4 \\
12 49 26.69 & -41 27 46.50 &  13.43 &  4866 &  113 &      &      &  4831 &      &  6 \\
12 49 30.20 & -41 25 15.70 &  17.63 &  2827 &   61 &      &      &  2818 &   30 &  1 \\
12 49 37.85 & -41 23 17.70 &  13.72 &  3628 &   24 &      &      &  3627 &   24 &  4 \\
12 49 40.14 & -41 21 58.30 &  16.01 &  3107 &  131 &      &      &  2880 &   40 &  1 \\
12 49 41.96 & -41 13 45.50 &  16.17 &  3657 &   69 &      &      &  3715 &   26 &  1 \\
12 49 51.56 & -41 13 34.70 &  13.69 &  2202 &   30 &      &      &  2160 &   23 &  1 \\
12 49 54.18 & -41 16 45.50 &  13.50 &  3822 &   25 &      &      &  3862 &    4 &  5 \\
12 49 56.00 & -41 24 04.40 &  17.33 &  4640 &   83 &      &      &  4661 &   69 &  1 \\
12 50 03.88 & -41 22 55.10 &  11.32 &  4614 &   28 &      &      &  4678 &    4 &  5 \\
12 50 11.53 & -41 13 15.80 &  14.12 &  2928 &   25 &      &      &  2908 &   26 &  1 \\
12 50 11.86 & -41 17 56.90 &  15.45 &  4105 &   49 &      &      &  4172 &   33 &  8 \\
12 50 12.15 & -41 30 56.00 &  13.00 &  2458 &   29 &      &      &  2384 &   24 &  1 \\
12 50 22.00 & -41 23 37.00 &  16.85 &  4450 &   77 &      &      &       &      &    \\
12 50 33.51 & -41 20 09.80 &  16.31 &  2278 &   95 &      &      &       &      &    \\
12 50 34.37 & -41 28 15.10 &  13.38 &  5316 &   31 &      &      &  5302 &    6 &  5 \\
12 50 34.63 & -41 27 07.10 &  17.30 &  4608 &  128 &      &      &       &      &    \\
12 50 57.45 & -41 23 47.60 &  14.34 &  4450 &   37 &      &      &  4445 &   19 &  1 \\
12 50 58.90 & -41 29 40.20 &  17.39 &  2112 &   87 &      &      &  2139 &   36 &  1 \\
12 51 00.86 & -41 43 21.10 &  13.37 &  2521 &   34 &      &      &  2469 &   23 &  1 \\
12 51 08.89 & -41 40 12.30 &  16.25 &  2646 &   85 &      &      &       &      &    \\
12 51 32.84 & -41 13 39.90 &  13.99 &  4775 &   35 &      &      &  4771 &   24 &  1 \\
12 51 36.34 & -41 29 32.50 &  15.13 &  3669 &   70 &      &      &  3864 &   23 &  1 \\
12 51 37.31 & -41 18 12.80 &  13.91 &  3502 &   33 &      &      &  3445 &   22 &  1 \\
12 51 39.78 & -41 28 55.60 &  17.14 &  5021 &   23 &      &      &  4879 &   58 &  1 \\
12 51 47.95 & -40 59 37.50 &  14.68 &  2657 &   29 &      &      &  2722 &   47 &  1 \\
12 51 50.83 & -41 11 10.60 &  15.23 &  4736 &   37 &      &      &  4712 &   18 &  1 \\
12 51 51.30 & -41 25 56.50 &  16.41 &  3560 &   60 &      &      &  3624 &   22 &  1 \\
12 51 56.43 & -41 32 20.70 &  13.86 &  3695 &   29 &      &      &  3682 &   19 &  1 \\
12 52 02.09 & -41 25 07.60 &  16.97 &  3075 &   64 &      &      &  3329 &   76 &  1 \\
12 52 02.38 & -41 21 11.70 &  17.06 &  4327 &   68 &      &      &       &      &    \\
12 52 03.08 & -41 27 34.60 &  13.89 &  4000 &   22 &      &      &  3945 &    3 &  5 \\
12 52 12.97 & -41 20 20.40 &  14.34 &  5134 &   80 &      &      &  5107 &   31 &  1 \\
12 52 15.47 & -41 28 37.20 &  15.49 &  3119 &   72 &      &      &  3037 &   26 &  1 \\
12 52 15.69 & -41 15 33.60 &  14.62 &  3476 &   52 &      &      &  3388 &   23 &  1 \\
12 52 16.02 & -41 23 26.80 &  13.31 &  3015 &   20 &      &      &  2984 &   23 &  4 \\
12 52 19.55 & -41 03 36.60 &  12.34 &  3381 &   36 &      &      &  3389 &   26 &  1 \\
12 52 21.52 & -41 10 00.60 &  17.24 &       &      & 4881 &   23 &  4881 &   45 &  1 \\
12 52 22.61 & -41 16 55.81 &  13.94 &  4627 &   27 &      &      &  4611 &      &  1 \\
12 52 30.41 & -40 55 43.10 &  15.69 &       &      & 2853 &   37 &       &      &    \\
12 52 32.86 & -41 09 44.00 &  17.81 &       &      & 4553 &   44 &       &      &    \\
12 52 34.82 & -41 13 33.60 &  17.40 &  4801 &   39 &      &      &  4971 &   71 &  1 \\
12 52 35.13 & -41 17 23.80 &  18.11 &       &      & 5059 &  113 &       &      &    \\
12 52 40.91 & -41 13 47.20 &  14.71 &  4764 &   41 &      &      &  4858 &   23 &  1 \\
12 52 42.58 & -41 09 57.70 &  15.66 &  4333 &   36 &      &      &       &      &    \\
12 52 50.18 & -41 20 14.70 &  12.92 &  4798 &   27 &      &      &  4868 &   45 &  7 \\
12 53 14.03 & -41 08 27.50 &  16.45 &  2845 &  114 &      &      &  2766 &   43 &  1 \\
12 53 20.19 & -41 38 07.50 &  13.61 &  4829 &   32 & 4840 &   50 &  4796 &   25 &  9 \\

\enddata

\tablenotetext{a}{Total magnitudes from SExtractor\citep{ba96}}

\tablenotetext{b}{references: (1)\citet{sjf97} (2)\citet{dcl86}
(3)\citet{mf96} (4)\citet{s96} (5)\citet{slhsd00} (6)\citet{lv89}
(7)\citet{dvdv91} (8)\citet{badwwprm02} (9)\citet{asp95} }

\end{deluxetable}


%% file: tab3.tex



\begin{deluxetable}{llllllllll}
\tabletypesize{\scriptsize}
\tablewidth{0pt}
\tablecaption{Background Galaxies\label{tablenon}}
\tablehead{
\colhead{$\alpha$} &
\colhead{$\delta$}      & 
\colhead{$V$} & 
\colhead{$v_{rad}$} &
\colhead{Error} &
\colhead{$v_{rad}$} & 
\colhead{Error} &
\colhead{$v_{rad}$} &
\colhead{Error} &
\colhead{reference\tablenotemark{b}} \\
\colhead{2000.0} &
\colhead{} &
\colhead{total\tablenotemark{a}} &
\colhead{Absorption} &
\colhead{} &
\colhead{Emission} &
\colhead{} &
\colhead{Literature} &
\colhead{} &
\colhead{} 
}

\startdata

11 47 28.96 & -41 17 02.10 & 18.60  & 11485 &  204 &       &      &       &     &    \\
12 47 35.07 & -41 10 46.90 & 19.51  & 55887 &   72 &       &      &       &      &    \\
12 47 36.50 & -41 21 36.20 & 18.15  & 38448 &   65 & 38510 &      &       &      &    \\
12 47 39.15 & -41 22 08.60 & 19.19  & 38723 &  131 & 38416 &   92 &       &      &    \\
12 47 41.18 & -41 13 24.30 & 18.78  & 63146 &   54 & 63291 &   85 &       &      &    \\
12 47 43.60 & -41 10 10.90 & 16.11  & 55845 &   63 & 55891 &      &       &      &    \\
12 47 47.34 & -41 36 07.70 & 17.63  &       &      & 31845 &   81 &       &      &    \\
12 47 47.44 & -41 37 28.70 & 19.55  & 70386 &  104 &       &      &       &      &    \\
12 47 48.39 & -41 22 38.40 & 17.92  & 38400 &   89 & 38453 &      &       &      &    \\
12 47 50.02 & -41 16 33.70 & 18.89  & 45664 &   94 & 46003 &      &       &      &    \\
12 47 52.07 & -41 12 01.50 & 19.18  & 70459 &  107 &       &      &       &      &    \\
12 47 54.18 & -41 38 59.00 & 17.50  & 25994 &   67 &       &      &       &      &    \\
12 47 55.98 & -41 00 24.70 & 19.43  & 29069 &   64 &       &      &       &      &    \\
12 47 59.11 & -41 25 30.10 & 18.15  & 40275 &   75 &       &      &       &      &    \\
12 48 01.58 & -41 26 52.80 & 18.43  & 70446 &  111 & 70356 &      &       &      &    \\
12 48 01.86 & -40 54 41.20 & 18.54  & 55424 &   73 &       &      &       &      &    \\
12 48 05.72 & -41 36 35.70 & 19.36  & 71829 &   97 &       &      &       &      &    \\
12 48 09.69 & -41 12 20.60 & 18.08  & 55980 &   57 &       &      &       &      &    \\
12 48 14.83 & -41 23 35.90 & 19.40  & 70420 &  140 & 70160 &      &       &      &    \\
12 48 15.63 & -40 59 40.50 & 19.00  & 45771 &  114 & 45945 &   37 &       &      &    \\
12 48 17.54 & -40 59 49.70 & 18.41  & 55607 &   84 &       &      &       &      &    \\
12 48 18.88 & -41 38 22.70 & 18.32  & 45831 &   83 &       &      &       &      &    \\
12 48 19.57 & -40 58 50.30 & 18.87  & 56054 &  113 &       &      &       &      &    \\
12 48 22.31 & -41 11 33.30 & 18.88  & 26129 &   76 & 25902 &   18 &       &      &    \\
12 48 24.25 & -41 13 16.80 & 19.17  & 55296 &  113 &       &      &       &      &    \\
12 48 25.34 & -41 26 07.00 & 18.75  & 56146 &   84 & 55913 &      &       &      &    \\
12 48 27.94 & -41 09 02.70 & 17.23  & 33594 &  102 &       &      & 33639 &   69 &  4 \\
12 48 28.17 & -41 21 46.00 & 18.35  & 49722 &  148 & 46740 &      &       &      &    \\
12 48 28.62 & -40 59 07.00 & 18.83  & 51793 &  106 & 51755 &      &       &      &    \\
12 48 29.19 & -41 27 21.20 & 17.88  & 46633 &   78 &       &      &       &      &    \\
12 48 32.84 & -41 15 35.00 & 18.28  & 38544 &  115 & 38509 &      &       &      &    \\
12 48 36.46 & -41 32 30.00 & 19.35  & 55295 &   89 &       &      &       &      &    \\
12 48 37.36 & -41 19 15.30 & 18.07  & 33146 &   86 & 33121 &   48 &       &      &    \\
12 48 37.98 & -41 27 10.20 & 18.04  & 39754 &   83 & 39720 &      &       &      &    \\
12 48 38.41 & -41 20 28.90 & 18.52  & 22032 &  107 & 25631 &      &       &      &    \\
12 48 38.77 & -41 07 05.50 & 17.56  & 38504 &   83 &       &      &       &      &    \\
12 48 39.71 & -41 12 13.10 & 18.40  & 56277 &  142 &       &      &       &      &    \\
12 48 40.49 & -41 28 32.70 & 18.95  & 46565 &  144 & 46766 &      &       &      &    \\
12 48 42.12 & -41 26 50.90 & 18.47  & 46457 &   99 & 46554 &      &       &      &    \\
12 48 43.51 & -41 28 21.00 & 19.75  & 72840 &  106 & 72243 &      &       &      &    \\
12 48 43.95 & -41 34 03.10 & 18.57  & 32933 &   88 &       &      &       &      &    \\
12 48 46.81 & -41 43 35.40 & 18.44  &       &      & 60896 &   55 &       &      &    \\
12 48 47.25 & -41 33 21.60 & 19.23  & 55105 &  108 & 55136 &      &       &      &    \\
12 48 48.57 & -41 16 16.60 & 17.35  & 56008 &   83 &       &      &       &      &    \\
12 48 52.40 & -41 44 50.20 & 18.49  & 69847 &   89 &       &      &       &      &    \\
12 48 54.22 & -41 30 12.20 & 18.03  & 39735 &  103 & 39739 &  266 &       &      &    \\
12 48 54.93 & -41 39 31.00 & 18.73  & 15900 &  118 &       &      &       &      &    \\
12 48 56.80 & -41 15 33.60 & 19.23  & 55885 &   70 &       &      &       &      &    \\
12 48 57.78 & -41 26 49.80 & 18.10  & 56103 &   80 &       &      &       &      &    \\
12 49 00.39 & -41 32 55.90 & 18.68  & 29027 &  107 & 29035 &      &       &      &    \\
12 49 00.88 & -41 25 42.20 & 17.76  & 16626 &   81 &       &      & 16514 &   64 &  1 \\
12 49 02.82 & -40 59 41.70 & 17.96  & 14769 &   35 &       &      &       &      &    \\
12 49 05.27 & -41 10 50.50 & 18.31  & 60715 &   66 &       &      &       &      &    \\
12 49 05.36 & -41 24 09.00 & 17.61  & 55683 &   62 &       &      &       &      &    \\
12 49 06.22 & -41 17 51.20 & 16.65  & 56099 &   83 &       &      &       &      &    \\
12 49 07.27 & -41 31 07.50 & 18.24  & 47355 &   56 &       &      &       &      &    \\
12 49 09.02 & -41 41 15.90 & 17.54  & 39564 &   80 &       &      &       &      &    \\
12 49 09.67 & -41 35 35.80 & 17.57  & 39596 &   73 &       &      &       &      &    \\
12 49 15.05 & -41 18 37.60 & 17.46  & 26286 &   45 &       &      &       &      &    \\
12 49 17.15 & -41 35 50.80 & 19.25  & 56310 &   97 &       &      &       &      &    \\
12 49 19.78 & -41 15 59.20 & 18.40  & 34244 &  130 & 34001 &      &       &      &    \\
12 49 20.53 & -41 38 46.50 & 19.02  &       &      & 14454 &   51 &       &      &    \\
12 49 20.70 & -41 19 17.02 & 17.39  & 25394 &   61 &       &      &       &      &    \\
12 49 24.37 & -41 36 59.90 & 19.00  & 56048 &   77 &       &      &       &      &    \\
12 49 25.59 & -41 30 24.90 & 17.09  &       &      & 10676 &   40 &       &      &    \\
12 49 28.66 & -41 34 31.60 & 18.54  & 71243 &   72 &       &      &       &      &    \\
12 49 30.27 & -41 22 34.90 & 19.43  & 55934 &   84 &       &      &       &      &    \\
12 49 30.39 & -41 18 18.00 & 19.21  &       &      & 55016 &  116 &       &      &    \\
12 49 31.39 & -41 13 40.20 & 18.51  &       &      &  9381 &   73 &       &      &    \\
12 49 36.56 & -41 16 41.90 & 18.99  & 12642 &  180 &       &      &       &      &    \\
12 49 41.54 & -41 41 50.40 & 19.18  &       &      & 22259 &   74 &       &      &    \\
12 49 48.99 & -41 39 38.90 & 19.69  & 64656 &   97 &       &      &       &      &    \\
12 49 51.91 & -41 13 58.30 & 18.36  & 13238 &  102 &       &      &       &      &    \\
12 49 59.16 & -41 36 16.40 & 17.95  & 71651 &   91 &       &      &       &      &    \\
12 50 00.85 & -41 43 26.50 & 16.79  & 16321 &   88 &       &      & 16362 &   45 &  1 \\
12 50 07.93 & -41 26 27.80 & 18.46  & 55911 &   68 &       &      &       &      &    \\
12 50 11.35 & -41 10 05.00 & 19.40  & 27181 &  102 &       &      &       &      &    \\
12 50 12.02 & -41 28 25.80 & 16.59  & 16234 &   53 &       &      &       &      &    \\
12 50 14.76 & -40 58 14.20 & 19.29  & 69788 &   93 &       &      &       &      &    \\
12 50 14.95 & -41 04 04.70 & 19.43  & 82514 &   80 &       &      &       &      &    \\
12 50 17.73 & -41 23 55.50 & 18.93  & 47318 &   84 &       &      &       &      &    \\
12 50 20.17 & -41 38 16.30 & 18.52  & 26698 &  110 &       &      &       &      &    \\
12 50 21.84 & -40 55 42.70 & 19.35  & 25980 &   90 & 25941 &   44 &       &      &    \\
12 50 27.31 & -41 26 00.20 & 17.65  & 44786 &  120 & 44650 &      &       &      &    \\
12 50 40.32 & -40 55 46.20 & 16.79  & 28935 &   64 &       &      &       &      &    \\
12 50 43.02 & -41 24 39.10 & 18.48  & 56375 &   87 &       &      &       &      &    \\
12 50 45.94 & -40 57 40.80 & 19.39  & 61826 &   93 &       &      &       &      &    \\
12 50 46.24 & -41 07 23.50 & 17.66  & 33438 &  166 & 33204 &      &       &      &    \\
12 50 55.08 & -40 58 11.40 & 18.46  &       &      & 76101 &  132 &       &      &    \\
12 50 55.72 & -41 21 46.90 & 18.58  &       &      & 56028 &  104 &       &      &    \\
12 50 57.69 & -41 02 06.60 & 18.39  & 56562 &   79 & 56760 &      &       &      &    \\
12 50 57.89 & -41 44 42.40 & 16.51  & 27458 &   68 &       &      &       &      &    \\
12 50 58.60 & -41 35 10.20 & 15.38  & 16376 &   61 &       &      & 16475 &      &  2 \\
12 51 02.78 & -41 26 47.50 & 18.45  &       &      & 10153 &   84 &       &      &    \\
12 51 03.57 & -41 08 01.40 & 18.42  &       &      & 13594 &   53 &       &      &    \\
12 51 07.60 & -41 26 03.90 & 18.20  & 56167 &  129 &       &      &       &      &    \\
12 51 07.80 & -41 19 46.10 & 19.10  & 46798 &   99 & 46334 &  605 &       &      &    \\
12 51 09.32 & -41 19 12.70 & 15.31  &       &      & 10653 &   33 & 10619 &      &  4 \\
12 51 12.55 & -41 10 57.80 & 18.34  &       &      & 10444 &   52 &       &      &    \\
12 51 15.57 & -41 34 25.60 & 17.93  & 21338 &  100 &       &      &       &      &    \\
12 51 17.72 & -41 16 34.10 & 17.69  &       &      & 10678 &   59 & 10679 &   38 &  1 \\
12 51 22.97 & -40 58 35.40 & 17.72  & 30732 &   72 &       &      &       &      &    \\
12 51 25.87 & -41 16 17.80 & 17.96  & 42571 &  103 &       &      &       &      &    \\
12 51 28.39 & -41 24 14.00 & 18.26  & 72641 &  118 &       &      &       &      &    \\
12 51 30.71 & -41 07 33.80 & 17.33  &       &      &  8726 &   52 &       &      &    \\
12 51 32.52 & -41 08 20.40 & 19.39  &  8620 &  148 &       &      &       &      &    \\
12 51 34.51 & -41 01 32.40 & 15.91  & 33970 &   49 &       &      &       &      &    \\
12 51 36.05 & -41 33 09.10 & 17.94  &  7578 &   73 &       &      &       &      &    \\
12 51 40.04 & -41 01 22.90 & 15.21  &       &      & 25588 &   75 &       &      &    \\
12 51 53.29 & -40 59 04.70 & 19.54  & 70608 &   96 &       &      &       &      &    \\
12 52 00.55 & -41 00 42.90 & 19.20  & 56396 &  106 & 56567 &      &       &      &    \\
12 52 04.10 & -41 30 01.00 & 17.79  & 27809 &   79 &       &      &       &      &    \\
12 52 05.03 & -41 37 39.80 & 18.74  & 19900 &  160 & 19706 &   21 &       &      &    \\
12 52 05.41 & -41 38 39.50 & 17.02  &       &      & 19413 &   70 & 19439 &   50 &  1 \\
12 52 05.41 & -41 33 48.50 & 17.26  & 27947 &   52 &       &      &       &      &    \\
12 52 06.09 & -41 22 27.30 & 18.14  & 67457 &   79 &       &      &       &      &    \\
12 52 06.31 & -41 37 35.40 & 17.58  & 19546 &   93 &       &      &       &      &    \\
12 52 08.02 & -41 17 57.50 & 18.52  & 45370 &  102 &       &      &       &      &    \\
12 52 13.53 & -41 40 50.40 & 18.80  & 27296 &   81 &       &      &       &      &    \\
12 52 19.57 & -40 57 06.90 & 18.70  & 70087 &  119 &       &      &       &      &    \\
12 52 23.66 & -41 21 28.90 & 16.89  & 25030 &   63 &       &      & 25120 &      &  4 \\
12 52 25.49 & -41 34 46.80 & 19.58  & 29541 &   70 & 29413 &      &       &      &    \\
12 52 25.91 & -41 29 18.10 & 19.12  &  6030 &  132 &       &      &       &      &    \\
12 52 27.34 & -41 03 36.60 & 18.62  & 56873 &  116 &       &      &       &      &    \\
12 52 28.73 & -41 26 05.40 & 16.00  & 31502 &   67 &       &      &       &      &    \\
12 52 32.22 & -41 36 06.30 & 17.61  & 27221 &   76 &       &      &       &      &    \\
12 52 32.86 & -41 32 19.50 & 16.95  &       &      & 29292 &   80 &       &      &    \\
12 52 33.30 & -41 11 05.20 & 18.60  &       &      & 44742 &   50 &       &      &    \\
12 52 36.05 & -41 15 33.90 & 17.61  & 23485 &   53 &       &      & 23457 &      &  4 \\
12 52 38.69 & -41 30 10.50 & 17.53  & 19121 &  129 &       &      &       &      &    \\
12 52 39.18 & -41 23 09.80 & 17.40  & 31479 &   57 &       &      & 31348 &   45 &  4 \\
12 52 41.13 & -41 15 13.60 & 17.37  & 56331 &   89 &       &      &       &      &    \\
12 52 41.56 & -40 58 24.40 & 17.98  & 56584 &   72 &       &      &       &      &    \\
12 52 43.50 & -41 18 43.40 & 18.13  & 25022 &   74 &       &      &       &      &    \\
12 52 45.32 & -41 28 00.20 & 16.40  & 24854 &   66 &       &      &       &      &    \\
12 52 46.99 & -41 19 36.30 & 17.01  & 56786 &   79 &       &      &       &      &    \\
12 52 50.67 & -41 00 46.00 & 18.66  & 37918 &  116 & 37717 &      &       &      &    \\
12 52 53.10 & -41 01 09.40 & 18.53  &       &      & 14629 &  109 &       &      &    \\
12 52 53.45 & -41 29 33.60 & 18.25  &  9331 &   63 &  9136 &   62 &       &      &    \\
12 52 55.06 & -41 12 28.60 & 18.67  &       &      & 19566 &   35 &       &      &    \\
12 52 57.21 & -41 22 57.40 & 18.17  & 56843 &   84 &       &      &       &      &    \\
12 52 59.42 & -41 32 40.90 & 18.35  &       &      & 60746 &  158 &       &      &    \\
12 53 05.35 & -41 42 39.90 & 15.85  & 27405 &   64 &       &      &       &      &    \\
12 53 17.38 & -41 18 27.90 & 16.30  & 27937 &   63 &       &      & 27974 &      &  4 \\
12 53 19.54 & -41 19 01.00 & 17.83  &  9343 &   66 &       &      &       &      &    \\
12 53 19.88 & -41 36 58.50 & 16.85  & 26216 &   81 &       &      &       &      &    \\
12 53 24.79 & -41 17 58.80 & 16.70  & 27900 &  104 &       &      & 27913 &   27 &  4 \\

\enddata

\tablenotetext{a}{Total magnitudes from SExtractor\citep{ba96}}

\tablenotetext{b}{references: (1)\citet{sjf97} (2)\citet{dcl86}
(3)\citet{mf96} (4)\citet{s96} (5)\citet{slhsd00} (6)\citet{lv89}
(7)\citet{dvdv91} (8)\citet{badwwprm02} (9)\citet{asp95} }

\end{deluxetable}


%% file: tab4.tex



\begin{deluxetable}{lllllll}
\tabletypesize{\scriptsize}
\tablewidth{0pt}
\tablecaption{New Radial Velocities\label{tablenew}}
\tablehead{
\colhead{$\alpha$} &
\colhead{$\delta$}      & 
\colhead{$V$} & 
\colhead{$v_{rad}$} &
\colhead{Error} &
\colhead{$v_{rad}$} & 
\colhead{Error} \\
\colhead{2000.0} &
\colhead{} &
\colhead{total\tablenotemark{a}} &
\colhead{(Absorption)} &
\colhead{} &
\colhead{(Emission)} &
\colhead{} 
}

\startdata

12 47 52.20 & -41 20 13.60 &  18.05 &   2828\tablenotemark{*} &   88 &      &     \\
12 47 54.18 & -41 03 44.90 &  15.56 &   4464 &   60 &      &     \\
12 47 55.84 & -40 54 16.50 &  17.33 &   1817\tablenotemark{*} &   74 &      &     \\
12 48 07.29 & -41 37 46.30 &  18.34 &   4188 &  106 &      &     \\
12 48 16.68 & -41 26 10.60 &  17.34 &   2118 &   90 &      &     \\
12 48 48.56 & -41 20 53.50 &  17.78 &   3053\tablenotemark{*} &   40 &      &     \\
12 50 22.00 & -41 23 37.00 &  16.85 &   4450 &   77 &      &     \\
12 50 33.51 & -41 20 09.80 &  16.31 &   2278 &   95 &      &     \\
12 50 34.63 & -41 27 07.10 &  17.30 &   4608 &  128 &      &     \\
12 51 08.89 & -41 40 12.30 &  16.25 &   2646 &   85 &      &    \\
12 52 02.38 & -41 21 11.70 &  17.06 &   4327 &   68 &      &     \\
12 52 30.41 & -40 55 43.10 &  15.69 &       &     &  2853\tablenotemark{*} &   37 \\
12 52 32.86 & -41 09 44.00 &  17.81 &       &     &  4553 &   44 \\
12 52 35.13 & -41 17 23.80 &  18.11 &       &     &  5059 &  113 \\
12 52 42.58 & -41 09 57.70 &  15.66 &   4333 &   36 &      &     \\

\enddata

\tablenotetext{a}{Total magnitudes from SExtractor\citep{ba96}}
\tablenotetext{*}{Galaxies are not included in the Centaurus Cluster
Catalog \citep{jd97}}

\end{deluxetable}


%% file: ms.bbl
\begin{thebibliography}{46}
\expandafter\ifx\csname natexlab\endcsname\relax\def\natexlab#1{#1}\fi

\bibitem[{{Adami} {et~al.}(1998){Adami}, {Nichol}, {Mazure}, {Durret},
  {Holden}, \& {Lobo}}]{anmdhl98}
{Adami}, C., {Nichol}, R.~C., {Mazure}, A., {Durret}, F., {Holden}, B., \&
  {Lobo}, C. 1998, \aap, 334, 765

\bibitem[{{Aguero} {et~al.}(1995){Aguero}, {Suarez}, \& {Paolantonio}}]{asp95}
{Aguero}, E.~L., {Suarez}, F., \& {Paolantonio}, S. 1995, \pasp, 107, 959

\bibitem[{{Bernardi} {et~al.}(2002){Bernardi}, {Alonso}, {da Costa}, {Willmer},
  {Wegner}, {Pellegrini}, {Rit{\' e}}, \& {Maia}}]{badwwprm02}
{Bernardi}, M., {Alonso}, M.~V., {da Costa}, L.~N., {Willmer}, C.~N.~A.,
  {Wegner}, G., {Pellegrini}, P.~S., {Rit{\' e}}, C., \& {Maia}, M.~A.~G. 2002,
  \aj, 123, 2990

\bibitem[{{Bernstein} {et~al.}(1995){Bernstein}, {Nichol}, {Tyson}, {Ulmer}, \&
  {Wittman}}]{bntuw95}
{Bernstein}, G.~M., {Nichol}, R.~C., {Tyson}, J.~A., {Ulmer}, M.~P., \&
  {Wittman}, D. 1995, \aj, 110, 1507+

\bibitem[{{Bertin} \& {Arnouts}(1996)}]{ba96}
{Bertin}, E. \& {Arnouts}, S. 1996, \aaps, 117, 393

\bibitem[{{Biviano} {et~al.}(1997){Biviano}, {Katgert}, {Mazure}, {Moles}, {den
  Hartog}, {Perea}, \& {Focardi}}]{bkmmhpf97}
{Biviano}, A., {Katgert}, P., {Mazure}, A., {Moles}, M., {den Hartog}, R.,
  {Perea}, J., \& {Focardi}, P. 1997, \aap, 321, 84

\bibitem[{{Chiboucas} \& {Mateo}(2006{\natexlab{a}})}]{KCMM1}
{Chiboucas}, K. \& {Mateo}, M. 2006{\natexlab{a}}, in preparation

\bibitem[{{Chiboucas} \& {Mateo}(2006{\natexlab{b}})}]{KCMM3}
---. 2006{\natexlab{b}}, in preparation

\bibitem[{{Conselice} {et~al.}(2001){Conselice}, {Gallagher}, \&
  {Wyse}}]{cgw01}
{Conselice}, C.~J., {Gallagher}, J.~S., \& {Wyse}, R.~F.~G. 2001, \apj, 559,
  791

\bibitem[{{De Propris} {et~al.}(2003){De Propris}, {Colless}, {Driver},
  {Couch}, {Peacock}, {Baldry}, {Baugh}, {Bland-Hawthorn}, {Bridges}, {Cannon},
  {Cole}, {Collins}, {Cross}, {Dalton}, {Efstathiou}, {Ellis}, {Frenk},
  {Glazebrook}, {Hawkins}, {Jackson}, {Lahav}, {Lewis}, {Lumsden}, {Maddox},
  {Madgwick}, {Norberg}, {Percival}, {Peterson}, {Sutherland}, \&
  {Taylor}}]{dep03}
{De Propris}, R., {Colless}, M., {Driver}, S.~P., {Couch}, W., {Peacock},
  J.~A., {Baldry}, I.~K., {Baugh}, C.~M., {Bland-Hawthorn}, J., {Bridges}, T.,
  {Cannon}, R., {Cole}, S., {Collins}, C., {Cross}, N., {Dalton}, G.~B.,
  {Efstathiou}, G., {Ellis}, R.~S., {Frenk}, C.~S., {Glazebrook}, K.,
  {Hawkins}, E., {Jackson}, C., {Lahav}, O., {Lewis}, I., {Lumsden}, S.,
  {Maddox}, S., {Madgwick}, D.~S., {Norberg}, P., {Percival}, W., {Peterson},
  B., {Sutherland}, W., \& {Taylor}, K. 2003, \mnras, 342, 725

\bibitem[{{de Propris} {et~al.}(1995){de Propris}, {Pritchet}, {Harris}, \&
  {McClure}}]{dphm95}
{de Propris}, R., {Pritchet}, C.~J., {Harris}, W.~E., \& {McClure}, R.~D. 1995,
  \apj, 450, 534+

\bibitem[{{de Vaucouleurs} {et~al.}(1991){de Vaucouleurs}, {de Vaucouleurs},
  {Corwin}, {Buta}, {Paturel}, \& {Fouque}}]{dvdv91}
{de Vaucouleurs}, G., {de Vaucouleurs}, A., {Corwin}, H.~G., {Buta}, R.~J.,
  {Paturel}, G., \& {Fouque}, P. 1991, {Third Reference Catalogue of Bright
  Galaxies} (Volume 1-3, XII, 2069 pp.~7 figs..~ Springer-Verlag Berlin
  Heidelberg New York)

\bibitem[{{Dickens} {et~al.}(1986){Dickens}, {Currie}, \& {Lucey}}]{dcl86}
{Dickens}, R.~J., {Currie}, M.~J., \& {Lucey}, J.~R. 1986, \mnras, 220, 679

\bibitem[{{Drinkwater} \& {Gregg}(1998)}]{dg98}
{Drinkwater}, M.~J. \& {Gregg}, M.~D. 1998, \mnras, 296, L15+

\bibitem[{{Drinkwater} {et~al.}(2000){Drinkwater}, {Jones}, {Gregg}, \&
  {Phillipps}}]{djgp00}
{Drinkwater}, M.~J., {Jones}, J.~B., {Gregg}, M.~D., \& {Phillipps}, S. 2000,
  Publications of the Astronomical Society of Australia, 17, 227

\bibitem[{{Driver} {et~al.}(1994){Driver}, {Phillipps}, {Davies}, {Morgan}, \&
  {Disney}}]{dpdmd94}
{Driver}, S.~P., {Phillipps}, S., {Davies}, J.~I., {Morgan}, I., \& {Disney},
  M.~J. 1994, \mnras, 268, 393+

\bibitem[{{Durret} {et~al.}(2002){Durret}, {Adami}, \& {Lobo}}]{dal02}
{Durret}, F., {Adami}, C., \& {Lobo}, C. 2002, \aap, 393, 439

\bibitem[{{Edwards} {et~al.}(2002){Edwards}, {Colless}, {Bridges}, {Carter},
  {Mobasher}, \& {Poggianti}}]{ecbcmp02}
{Edwards}, S.~A., {Colless}, M., {Bridges}, T.~J., {Carter}, D., {Mobasher},
  B., \& {Poggianti}, B.~M. 2002, \apj, 567, 178

\bibitem[{{Gavazzi} {et~al.}(2004){Gavazzi}, {Zaccardo}, {Sanvito}, {Boselli},
  \& {Bonfanti}}]{gzsbb04}
{Gavazzi}, G., {Zaccardo}, A., {Sanvito}, G., {Boselli}, A., \& {Bonfanti}, C.
  2004, \aap, 417, 499

\bibitem[{{Geha} {et~al.}(2003){Geha}, {Guhathakurta}, \& {van der
  Marel}}]{ggm03}
{Geha}, M., {Guhathakurta}, P., \& {van der Marel}, R.~P. 2003, \aj, 126, 1794

\bibitem[{{Giovanelli} {et~al.}(1982){Giovanelli}, {Haynes}, \&
  {Chincarini}}]{ghc82}
{Giovanelli}, R., {Haynes}, M.~P., \& {Chincarini}, G.~L. 1982, \apj, 262, 442

\bibitem[{{Hradecky} {et~al.}(2000){Hradecky}, {Jones}, {Donnelly},
  {Djorgovski}, {Gal}, \& {Odewahn}}]{hjddgo00}
{Hradecky}, V., {Jones}, C., {Donnelly}, R.~H., {Djorgovski}, S.~G., {Gal},
  R.~R., \& {Odewahn}, S.~C. 2000, \apj, 543, 521

\bibitem[{{Jerjen} \& {Dressler}(1997)}]{jd97}
{Jerjen}, H. \& {Dressler}, A. 1997, \aaps, 124, 1

\bibitem[{{Kurtz} \& {Mink}(1998)}]{km98}
{Kurtz}, M.~J. \& {Mink}, D.~J. 1998, \pasp, 110, 934

\bibitem[{{Landolt}(1992)}]{land92}
{Landolt}, A.~U. 1992, \aj, 104, 340

\bibitem[{{Lauberts} \& {Valentijn}(1989)}]{lv89}
{Lauberts}, A. \& {Valentijn}, E.~A. 1989, {The surface photometry catalogue of
  the ESO-Uppsala galaxies} (Garching: European Southern Observatory, |c1989)

\bibitem[{{Mathewson} \& {Ford}(1996)}]{mf96}
{Mathewson}, D.~S. \& {Ford}, V.~L. 1996, \apjs, 107, 97

\bibitem[{{Pedraz} {et~al.}(2002){Pedraz}, {Gorgas}, {Cardiel}, {S{\'
  a}nchez-Bl{\' a}zquez}, \& {Guzm{\' a}n}}]{pgcsg02}
{Pedraz}, S., {Gorgas}, J., {Cardiel}, N., {S{\' a}nchez-Bl{\' a}zquez}, P., \&
  {Guzm{\' a}n}, R. 2002, \mnras, 332, L59

\bibitem[{{Phillipps} {et~al.}(1998){Phillipps}, {Parker}, {Schwartzenberg}, \&
  {Jones}}]{ppsj98}
{Phillipps}, S., {Parker}, Q.~A., {Schwartzenberg}, J.~M., \& {Jones}, J.~B.
  1998, \apjl, 493, L59

\bibitem[{{Schechter}(1976)}]{s76}
{Schechter}, P. 1976, \apj, 203, 297

\bibitem[{{Secker} {et~al.}(1998){Secker}, {Harris}, {C\^{o}t\'{e}}, \&
  {Oke}}]{shco98}
{Secker}, J., {Harris}, W.~E., {C\^{o}t\'{e}}, P., \& {Oke}, J.~B. 1998, in
  Untangling Coma Berenices: A New Vision of an Old Cluster, 115--+

\bibitem[{{Secker} {et~al.}(1997){Secker}, {Harris}, \& {Plummer}}]{shp97}
{Secker}, J., {Harris}, W.~E., \& {Plummer}, J.~D. 1997, \pasp, 109, 1377

\bibitem[{{Smith} {et~al.}(2000){Smith}, {Lucey}, {Hudson}, {Schlegel}, \&
  {Davies}}]{slhsd00}
{Smith}, R.~J., {Lucey}, J.~R., {Hudson}, M.~J., {Schlegel}, D.~J., \&
  {Davies}, R.~L. 2000, \mnras, 313, 469

\bibitem[{{Smith} {et~al.}(1997){Smith}, {Driver}, \& {Phillipps}}]{sdp97}
{Smith}, R.~M., {Driver}, S.~P., \& {Phillipps}, S. 1997, \mnras, 287, 415

\bibitem[{{Stein}(1996)}]{s96}
{Stein}, P. 1996, \aaps, 116, 203

\bibitem[{{Stein} {et~al.}(1997){Stein}, {Jerjen}, \& {Federspiel}}]{sjf97}
{Stein}, P., {Jerjen}, H., \& {Federspiel}, M. 1997, \aap, 327, 952

\bibitem[{{Toft} {et~al.}(2004){Toft}, {Mainieri}, {Rosati}, {Lidman},
  {Demarco}, {Nonino}, \& {Stanford}}]{tmrldns04}
{Toft}, S., {Mainieri}, V., {Rosati}, P., {Lidman}, C., {Demarco}, R.,
  {Nonino}, M., \& {Stanford}, S.~A. 2004, \aap, 422, 29

\bibitem[{{Tonry} \& {Davis}(1979)}]{td79}
{Tonry}, J. \& {Davis}, M. 1979, \aj, 84, 1511

\bibitem[{{Trentham}(1997{\natexlab{a}})}]{t97b}
{Trentham}, N. 1997{\natexlab{a}}, \mnras, 290, 334

\bibitem[{{Trentham}(1997{\natexlab{b}})}]{t97}
---. 1997{\natexlab{b}}, \mnras, 286, 133

\bibitem[{{Trentham}(1998)}]{tcoma98}
---. 1998, \mnras, 293, 71

\bibitem[{{Trentham} {et~al.}(2001){Trentham}, {Tully}, \& {Verheijen}}]{ttv01}
{Trentham}, N., {Tully}, R.~B., \& {Verheijen}, M.~A.~W. 2001, \mnras, 325, 385

\bibitem[{{Trujillo} {et~al.}(2001){Trujillo}, {Aguerri}, {Cepa}, \& {Guti{\'
  e}rrez}}]{tacg01}
{Trujillo}, I., {Aguerri}, J.~A.~L., {Cepa}, J., \& {Guti{\' e}rrez}, C.~M.
  2001, \mnras, 328, 977

\bibitem[{{Valotto} {et~al.}(2001){Valotto}, {Moore}, \& {Lambas}}]{vml01}
{Valotto}, C.~A., {Moore}, B., \& {Lambas}, D.~G. 2001, \apj, 546, 157

\bibitem[{{Valotto} {et~al.}(2004){Valotto}, {Muriel}, {Moore}, \&
  {Lambas}}]{vmml03}
{Valotto}, C.~A., {Muriel}, H., {Moore}, B., \& {Lambas}, D.~G. 2004, \apj,
  603, 67

\bibitem[{{Valotto} {et~al.}(1997){Valotto}, {Nicotra}, {Muriel}, \&
  {Lambas}}]{vnml97}
{Valotto}, C.~A., {Nicotra}, M.~A., {Muriel}, H., \& {Lambas}, D.~G. 1997,
  \apj, 479, 90+

\bibitem[{{van Zee} {et~al.}(2004){van Zee}, {Skillman}, \& {Haynes}}]{zsh04}
{van Zee}, L., {Skillman}, E.~D., \& {Haynes}, M.~P. 2004, \aj, 128, 121

\end{thebibliography}
